\documentclass[12pt]{article}
\usepackage[affil-it]{authblk}

\usepackage[centertags]{amsmath}
\usepackage{url}
\usepackage{geometry}
\usepackage{hyperref}
\usepackage[latin1]{inputenc}
\usepackage{amsfonts,amssymb,amsmath,amscd,slashed,amsthm,mathtools,mathrsfs}
\usepackage{enumerate}
\usepackage{verbatim}

\usepackage[english]{babel}

\usepackage{graphicx}

\usepackage[all]{xy}
\usepackage{lmodern}

\usepackage{tikz}
\usepackage{tikz-cd}
\usetikzlibrary{shapes.misc,arrows,decorations.markings}
\usetikzlibrary{matrix}

\usepackage{color}
\usepackage{psfrag}
\usepackage{epsfig}
\usepackage{pdflscape}

\newtheorem{Definition}{Definition}
\newtheorem{Conjecture}[Definition]{Conjecture}
\newtheorem{Example}[Definition]{Example}
\newtheorem{Theorem}[Definition]{Theorem}
\newtheorem{Claim}[Definition]{Claim}
\newtheorem{Lemma}[Definition]{Lemma}
\newtheorem{Remark}[Definition]{Remark}
\newtheorem{Corollary}[Definition]{Corollary}
\newtheorem{Proposition}[Definition]{Proposition}

\newenvironment{pf}{\textbf{Proof}}{ \qed \\}
\newenvironment{theorem}[1][]{\begin{Theorem} \normalfont \textbf{#1} \itshape}{\end{Theorem}}

\newenvironment{definition}[1][]{\begin{Definition} \normalfont \textbf{#1} \itshape}{\end{Definition}}

\newenvironment{lemma}[1][]{\begin{Lemma} \normalfont \textbf{#1} \itshape}{\end{Lemma}}
\newenvironment{remark}[1][]{\begin{Remark} \normalfont \textbf{#1}}{\end{Remark}}
\newenvironment{corollary}[1][]{\begin{Corollary} \normalfont \textbf{#1}}{\end{Corollary}}
\newenvironment{proposition}[1][]{\begin{Proposition} \normalfont \textbf{#1}}{\end{Proposition}}

\newcommand{\sN}{{\mathbb N}}

\newcommand{\sR}{{\mathbb R}}
\newcommand{\sC}{{\mathbb C}}
\newcommand{\sM}{{\mathbb M}}

\newcommand{\Cf}{\mathscr{C}}
\newcommand{\E}{\mathcal{E}}
\newcommand{\F}{\mathcal{F}}
\newcommand{\I}{\mathcal{I}}
\newcommand{\J}{\mathcal{J}}
\newcommand{\M}{\mathcal{M}}
\newcommand{\Pf}{\mathscr{P}}
\newcommand{\K}{\mathcal{K}}
\newcommand{\T}{\mathcal{T}}
\newcommand{\cR}{\mathcal{R}}
\newcommand{\cS}{\mathcal{S}}
\newcommand{\X}{\mathcal{X}}
\newcommand{\U}{\mathcal{U}}
\newcommand{\Uf}{\mathscr{U}}

\newcommand{\itSigma}{\mathit{\Sigma}}
\newcommand{\itPi}{\mathit{\Pi}}

\newcommand{\ii}{\textrm{i}}

\newcommand{\bx}{\textup{\textbf{x}}}
\newcommand{\ba}{\mathbf{a}}
\newcommand{\bb}{\mathbf{b}}
\newcommand{\bv}{\mathbf{v}}
\newcommand{\bu}{\mathbf{u}}
\newcommand{\bw}{\mathbf{w}}
\newcommand{\bz}{\mathbf{z}}
\newcommand{\bupr}{\textup{\textbf{r}}}
\newcommand{\bp}{\boldsymbol{p}}
\newcommand{\bq}{\boldsymbol{q}}
\newcommand{\br}{\boldsymbol{r}}
\newcommand{\bitv}{\boldsymbol{v}}
\newcommand{\bgamma}{\boldsymbol{\gamma}}
\newcommand{\bmu}{\boldsymbol\mu}
\newcommand{\bnu}{\boldsymbol\nu}
\newcommand{\bomega}{\boldsymbol\omega}
\newcommand{\bsigma}{\boldsymbol\sigma}

\newcommand{\mN}{\scriptscriptstyle (N)}
\newcommand{\MN}{\M_{\mN}}

\newcommand{\jdn}{\textbf{1}}

\DeclareMathOperator{\id}{id}
\DeclareMathOperator{\supp}{supp}
\DeclareMathOperator{\ev}{ev}
\DeclareMathOperator{\im}{im}

\newcommand{\norm}[1]{\left\Vert {#1} \right\Vert}
\newcommand{\vc}{\vcentcolon =}
\newcommand{\cv}{= \vcentcolon}
\newcommand{\x}{\times}

\begin{document}


\title{Generally covariant $N$-particle dynamics}
\author{Tomasz Miller${}^{1}$\thanks{Corresponding author: tomasz.miller@uj.edu.pl}}
\author{Micha{\l} Eckstein${}^{2,3,1}$}
\author{Pawe{\l} Horodecki${}^{4,5}$}
\author{Ryszard Horodecki${}^{4}$}

\affil{\small ${}^1$ Copernicus Center for Interdisciplinary Studies, Jagiellonian University,
\\Szczepa\'nska 1/5, 31-011 Krak\'ow, Poland}
\affil{\small ${}^2$ Institute of Theoretical Physics, Jagiellonian University,
\\Profesora Stanis{\l}awa {\L}ojasiewicza 11, 30-348 Krak\'ow, Poland}
\affil{\small ${}^3$ Institute of Theoretical Physics and Astrophysics,
National Quantum Information Centre, 
\\ Faculty of Mathematics, Physics and Informatics, University  of  Gda\'nsk, \\ Wita  Stwosza  57,  80-308  Gda\'nsk,  Poland}
\affil{\small ${}^4$ International Centre for Theory of Quantum Technologies, University of Gda\'nsk, \\Wita Stwosza 63, 80-308 Gda\'nsk, Poland}
\affil{\small ${}^5$ Faculty of Applied Physics and Mathematics, National Quantum Information Centre,\\
Gda\'nsk University of Technology, Gabriela Narutowicza 11/12, 80-233 Gda\'nsk, Poland}

\maketitle

\begin{abstract}
A simultaneous description of the dynamics of multiple particles requires a configuration space approach with an external time parameter. This is in stark contrast with the relativistic paradigm, where time is but a coordinate chosen by an observer. Here we show, however, that the two attitudes toward modelling $N$-particle dynamics can be conciliated within a generally covariant framework. To this end we construct an `$N$-particle configuration spacetime' $\MN$, starting from a globally hyperbolic spacetime $\M$ with a chosen smooth splitting into time and space components. The dynamics of multi-particle systems is modelled at the level of Borel probability measures over $\MN$ with the help of the global time parameter. We prove that with any time-evolution of measures, which respects the $N$-particle causal structure of $\MN$, one can associate a single measure on the Polish space of `$N$-particle wordlines'. The latter is a splitting-independent object, from which one can extract the evolution of measures for any other global observer on $\M$. An additional asset of the adopted measure-theoretic framework is the possibility to model the dynamics of indistinguishable entities, such as quantum particles. As an application we show that the multi-photon and multi-fermion Schr\"odinger equations, although explicitly dependent on the choice of an external time-parameter, are in fact fully compatible with the causal structure of the Minkowski spacetime.
\end{abstract}

MSC classes: 53C50, 53C80, 28E99, 60B05


\section{Introduction}\label{sec:intro}
In modern physics there exist two approaches to modelling the dynamics of physical systems. The first one assumes a covariant point of view, based on the concept of a relativistic spacetime. In this context, the system is modelled by an entity, which does not evolve per se, e.g., a world-line or a field configuration. Time is but a coordinate associated with the choice of a (local or global) observer. The second approach exploits the concept of a configuration space (or, more generally, a phase space). From this perspective time is an external parameter bearing no relationship to the space itself.

These two viewpoints are not easily conciliated when the studied system involves multiple particles (or ``constituents''). Indeed, if one starts from a relativistic standpoint one firstly needs to choose a (local) coordinate chart or a (global) time-foliation. 
Not only such a choice is not canonical (even in the single-particle setting), but also it is not at all clear how to ascribe spatial coordinates to different particles at a single time-instance. On the other hand, within the configuration space approach one is free to include as many (generalised) coordinates as needed, while keeping a single external time parameter. However, such a framework heavily depends upon the observer and there is no canonical way to compare the descriptions of the same system by two observers adopting different time parameters.

In this paper we build a bridge between the two approaches to dynamics. We start with a globally hyperbolic spacetime $\M$ with a chosen splitting $\M \cong \sR \times \itSigma$ and we construct the `$N$-particle configuration spacetime' as $\MN \vc \sR \times \itSigma^N$. It is endowed with a causal structure pulled back from $\M$, which encodes the demand that the speed of every single particle must be bounded by $c$. Then, we employ the measure-theoretic formalism developed in our previous works \cite{PRA2017,AHP2017,Miller17} and show that the `$N$-particle causal order' admits a natural extension to the space $\Pf(\MN)$ of Borel probability measures over $\MN$. In this context, we introduce an evolution of measures as a certain map $\sR \supset I \ni t \mapsto \bmu_t \in \Pf(\MN)$ and discuss its compatibility with the causal structure. Although the entire construction seems to depend upon the splitting $\M \cong \sR \times \itSigma$ adopted at the very beginning, we show that general covariance is restored at the level of $N$-particle unparametrised causal curves. Concretely, we show (Theorem \ref{NcurvesMain}) that the causality of time-evolution of measures is equivalent to the existence of a single measure on the Polish space of `$N$-particle worldlines' (Definition \ref{NcurvesDef}), which is a splitting-independent object.

The (a)causality of joint dynamics of $N$ classical particles, determined by some equation specifying their trajectories $t \mapsto x_i(t) \in \itSigma$, ($i = 1, \ldots, N$) can be studied through an evolution of a delta-like measure $\bmu_t = \delta_t \times \delta_{x_1(t)} \times \delta_{x_2(t)} \times \cdots \times \delta_{x_N(t)}$. But the measure-theoretic framework is much more capacious. It allows one to model the dynamics of normalised statistical ensembles, such as dust density or charge distribution, but also quantum probability densities derived within the wave packet formalism \cite{PRA2020,AHP2017}. A new distinctive element available in the multi-particle setting are the correlations among particles. In particular, symmetric measures, e.g., $\delta_t \times \tfrac{1}{2} \big( \mu_t \times \nu_t + \nu_t \times \mu_t \big)$, are suited to model indistinguishable particles. On this occasion, let us emphasise that we use the term `particle' in a broad sense, not limited to neither classical nor quantum theory --- see \cite{PRA2020} for a detailed discussion.

Another gain from the measure-theoretic approach is its close relationship to the Lorentzian optimal transport theory. The latter is a new and fast-developing area of research \cite{Bertrand2013,Brenier2003,CavalettiMondino20,Kell2018,Suhr2016}, which has found successful applications in the early universe reconstruction problem \cite{BrenierFrisch2003,Frisch2002,Frisch2011,Frisch2006} and, more recently, in the studies linking general relativity and the second law of thermodynamics \cite{McCann18,Suhr18}.

The pertinence of integrating the configuration and spacetime perspectives to dynamics is manifest in the context of (quantum) information theory. On the one hand, information protocols involving multiple parties and multiple signals are described from an external perspective. On the other hand, the admissible communication schemes are severely constrained by the spacetime structure \cite{QIandGR}.

As an application of the developed formalism we inspect two multi-particle Schr\"odinger equations: ``multi-photon'' and ``multi-fermion'' equations. The former is a valuable concept in quantum optics \cite{Multiphoton}, utilised e.g. to study decoherence upon propagation of photons through turbulent atmosphere \cite{2photon}. The latter is a free (i.e. non-interacting) variant of the multi-particle Dirac operator, which is the basis for Dirac--Fock equations developed and applied in the domain of  quantum atomic physics and chemistry \cite{Dyall2007,Esteban1999,FroeseFischer2016,Grant2007,Levitt2014}.

We show that both multi-photon and multi-fermion equations, which belong to the configuration space realm, are fully compatible with the structure of relativistic spacetime in the rigorous sense of condition \eqref{Ncausal_evo}. To this end, we prove a more general result. Namely, we show that a measure enjoying the $N$-particle continuity equation (cf., for instance, \cite{Bernard12} and \cite{Crippa2007PhD}) with a subluminal multi-velocity field evolves causally in a well-defined covariant sense. This result generalises previous findings on the relationship between the continuity equation and the causality of the evolution of quantum probability densities \cite{PRA2017,Gerlach1969,Gerlach1968,Gromes1970}. It is worth emphasising that, although technically straightforward, this generalisation is far-reaching, as it incorporates the phenomenon of entanglement between quantum particles.

The plan of the paper is as follows: In Section \ref{sec:relation}, after recollecting some rudiments of causality theory, we introduce the `$N$-particle configuration spacetime' and uncover the basic features of its causal structure. Then, in Section \ref{sec:NJplus} we move on to the measure-theoretic realm. We show that the space $\Pf(\MN)$ is equipped with a natural causal order inherited from $\MN$ and provide its several equivalent characterisations (Theorem \ref{thm::equiv}). Section \ref{sec:evo} begins with a discussion of various Polish spaces of $N$-particle causal curves and concludes with the announced restoration of observer-independence (Theorem \ref{NcurvesMain}). Section \ref{sec:wave_eq} is devoted to the study of causality of multi-particle Schr\"odinger equations mentioned above. We finish with a brief summary and outlook into some future prospects.


\section{\texorpdfstring{$N$-particle configuration spacetime}{N-particle configuration spacetime}}\label{sec:relation}

\subsection{Preliminaries: elements of causality theory}\label{subsec:one-particle}

In order to fix the notation and make the article self-contained, we begin with a brief recollection of the elements of causality theory. For the full story the Reader is referred e.g. to \cite{Beem,MS08,BN83,Penrose1972}.

Let $\M$ be a spacetime, i.e. an $n$-dimensional connected time-oriented smooth Lorentzian manifold\footnote{We adopt the signature convention $(-++\ldots+)$.}. The Lorentzian metric on $\M$ induces binary relations $\ll$ and $\preceq$ on $\M$, called the \emph{chronological} and the \emph{causal precedence relations}, respectively. We say that $p$ chronologically (reps. causally) precedes $q$, or that an event $q$ is in the chronological (resp. causal) future of $p$, which is denoted $p \ll q$ (resp. $p \preceq q$), if there exists a piecewise smooth future-directed chronological (resp. causal) curve from $p$ to $q$. By the standard convention, we also assume $p \preceq p$. By $p \prec q$ we mean that $p \preceq q$, but $p \neq q$.

The relation $\preceq$ allows to extend the notion of a causal curve to the curves which are only $C^0$. Although in general the notion of a continuous (future-directed) causal curve is somewhat convoluted (cf. \cite[Definition 3.15]{MS08}), for the so-called distinguishing spacetimes it simplifies greatly. Namely \cite[Proposition 3.19]{MS08}, the map $\gamma \in C(I,\M)$ (i.e. a continuous map from an interval $I \subset \sR$ to $\M$) is called future-directed causal if, for any $s,t \in I$, $s < t$ implies that $\gamma(s) \prec \gamma(t)$.

Even though from the mathematical side the relations $\ll$ and $\preceq$ are subsets of $\M^2$, for historical reasons one usually denotes these subsets as $I^+ \vc \{ (p,q) \in \M^2 \ | \ p \ll q \}$ and $J^+ \vc \{ (p,q) \in \M^2 \ | \ p \preceq q \}$. One also writes $I^\pm(p)$ (resp. $J^\pm(p)$) to denote the set of all events in the chronological (resp. causal) future/past of $p$. Moreover, for any $\X \subset \M$ one introduces $I^\pm(\X) \vc \bigcup_{p \in \X} I^\pm(p)$ and similarly $J^\pm(\X) \vc \bigcup_{p \in \X} J^\pm(p)$. Finally, $I^\pm(p,U)$ denotes the set of events that can be reached from $p$ by means of future/past-directed timelike curves with images contained in $U \subset \M$.

A function $\T: \M \rightarrow \sR$ is referred to as
\begin{itemize}
\item a \emph{causal} function if $p \preceq q$ implies $\T(p) \leq \T(q)$,
\item a \emph{time} function if it is continuous and $p \prec q$ implies $\T(p) < \T(q)$,
\item a \emph{temporal} function if it is smooth and has a past-directed timelike gradient.
\end{itemize}
Every temporal function is a time function, but a smooth time function need not be temporal.

A \emph{Cauchy hypersurface} is a subset $\cS \subset \M$ met exactly once by every inextendible timelike curve. Any such $\cS$ is achronal (i.e. $\forall p,q \in \cS \ p \not\ll q$ or, equivalently, $I^+(\cS) \cap I^-(\cS) = \emptyset$) and it can be proven to be a closed topological hypersurface \cite{BN83} (see also Proposition \ref{Ncausality_Prop3} below). Time and temporal functions are called \emph{Cauchy} if all their level sets happen to be Cauchy hypersurfaces.

A spacetime $\M$ is called \emph{causal} if it does not contain causal loops, what happens if and only if $\preceq$ is a partial order. A more refined causal structure arises in \emph{globally hyperbolic spacetimes}, which, in addition to being causal, have the property that the intersections $J^+(p) \cap J^-(q)$ are compact for every $p,q \in \M$. A spacetime is globally hyperbolic iff it admits Cauchy temporal functions, which has a further striking consequence proven by Bernal and S\'anchez \cite{BS04}, who strengthened the earlier seminal result by Geroch \cite{GerochSplitting}.
\begin{theorem}[(Geroch--Bernal--S\'anchez)]
\label{GBSthm}
Let $\M$ be a globally hyperbolic spacetime with metric $g$. $\M$ admits Cauchy temporal functions, and for any such function $\T$ there exists an isometry $\Phi: \M \rightarrow \sR \times \itSigma$, which we shall call the Geroch--Bernal--S\'anchez (GBS) splitting, such that $\itSigma \vc \T^{-1}(0)$, $\T = \Phi^\ast \pi^0$ and the metric splits into
\begin{align*}
g = -\alpha \, d\T \otimes d\T + \bar{g},
\end{align*}
where $\alpha:\M \rightarrow \sR$ is a positive smooth function and $\bar{g}$ is a 2-covariant symmetric tensor field on $\M$ whose restriction to $\Phi^{-1}(\{t\} \times \itSigma) = \T^{-1}(t)$ is a Riemannian metric for every $t \in \sR$ and whose radical at each $p\in\M$ is spanned by the gradient $(\nabla \T)_p$.
\end{theorem}
Intuitively speaking, $\Phi$ splits $\M$ into the time and space parts, with the notion of time prescribed by $\T$ chosen by the observer. As we shall see below, Theorem 1 is the cornerstone of the proposed generalisation of causality theory to multi-particle systems.


\subsection{\texorpdfstring{$N$-particle causality theory}{N-particle causality theory}}\label{subsec:multi-particle}

In order to develop the $N$-particle extension of the standard causality theory, we first have to establish a suitable $N$-particle counterpart of a spacetime. At the first glance, a natural candidate seems to be simply $\M^N$ --- the $N$-th Cartesian power of $\M$. This, however, would result in a multitude of time coordinates in every chart with each time coordinate corresponding to one of the particles. Meanwhile, a point of an `$N$-particle configuration spacetime' should rather correspond to a spatial configuration of the $N$-particle system at a given time instant. This, of course, immediately raises questions regarding the observer independence of any such structure. Nevertheless, this seems to be an inevitable starting point. 

Bearing the above in mind, let $\M$ be a globally hyperbolic spacetime and let us fix a Cauchy temporal function $\T$. The latter determines the GBS splitting $\Phi: \M \rightarrow \sR \times \itSigma$ as given by Theorem \ref{GBSthm}. Now, by the `$N$-particle configuration spacetime' (associated to $\T$) we shall understand the product manifold $\MN \vc \sR \times \itSigma^N$. In order to understand its relation to $\M^N$, let us introduce the embedding $\iota: \MN \hookrightarrow \M^N$ defined as
\begin{align*}
\forall \, (t,x_1,\ldots,x_N) \in \MN \quad \ \iota(t,x_1,\ldots,x_N) \vc \left( \Phi^{-1}(t,x_1), \ldots, \Phi^{-1}(t,x_N) \right).
\end{align*}
The image of our `$N$-particle configuration spacetime' under this embedding is
\begin{align*}
\iota (\MN) = \left\{ (p_1,\ldots,p_N) \in \M^N | \ \T(p_1) = \ldots = \T(p_N) \right\}.
\end{align*}
Additionally, let us define $\iota^j: \MN \rightarrow \M$ via $\iota^j = \pi^j \circ \iota$, where $\pi^j$ is the canonical projection on the $j$-th spatial argument $j=1,\ldots,N$, and we shall also be using $\pi^0$ to denote the projection on the time component. The maps $\iota^j$ are clearly submersions, and with their help we are able to endow $\MN$ with a causal structure by pulling back the causal relations from $\M$. The interplay between the above discussed manifolds is summarised in the following commutative diagram
\begin{center}
\begin{tikzcd}
\MN \arrow[bend right=80, twoheadrightarrow, swap]{dd}{\pi^0} \arrow[twoheadrightarrow]{d}[swap]{(\pi^0,\pi^j)} \arrow[hookrightarrow]{r}{\iota} \arrow[twoheadrightarrow]{rd}{\iota^j} & \M^N \arrow[twoheadrightarrow]{d}{\pi^j}
\\
\sR \times \itSigma \arrow[twoheadrightarrow]{d}[swap]{\pi^0} \arrow[shift left]{r}{} & \M \arrow[shift left]{l}{\Phi} \arrow[twoheadrightarrow]{ld}{\T}
\\
\sR & 
\end{tikzcd}
\end{center}
For convenience, let us introduce the following notation. Objects living in $\MN$ will be denoted in bold italics, e.g., $\bp \vc (t,x_1,\ldots,x_N) \in \MN$. However, if they appear with a superscript, this signifies that they have been transported by means of $\iota^j$ and live in $\M$, i.e. $\bp^j \vc \iota^j(\bp) = \Phi^{-1}(t,x_j)$. 

Furthermore, for any $t \in \sR$ it will prove convenient to denote $\itSigma^N_t \vc \{t\} \times \itSigma^N$ and $\itSigma_t \vc \{t\} \times \itSigma$. Let us warn that $\itSigma^N_t$ is not the $N$-th Cartesian power of $\itSigma_t$, but since we will never need the latter, it should not lead to confusion.

As an immediate consequence of the fact that $\iota^j = \Phi^{-1} \circ (\pi^0, \pi^j)$, we have the following observation
\begin{proposition}
\label{Ncausality_Prop0}
The map $\iota^j$ is open. Moreover, the sequence $(\bp_k) \subset \MN$ converges to $\bp \in \MN$ iff the sequences $(\bp_k^j) \subset \M$ converge to $\bp^j \in \M$ for every $j=1,\ldots,N$.
\end{proposition}

As announced above, we now define the basic notions of causality theory in the $N$-particle setting by pulling them back from $\M$ with the aid of $\iota^j$'s. Let us emphasise that we do not endow $\MN$ with the structure of a Lorentzian manifold, i.e. we do not introduce any Lorentzian metric on $\MN$. What we ``borrow'' from the underlying globally hyperbolic spacetime $\M$ is just the causal structure.
\begin{definition}
\label{Ncausality_Def}
\
\begin{enumerate}[(i)]
\item For any $\bp \in \MN$, a tangent vector $\bitv \in T_{\bp}\MN$ is (future-directed) timelike (resp. causal) if the vectors $\bitv^j \vc \textrm{d}\iota^j(\bitv) \in T_{\bp^j}\M$ are (future-directed) timelike (resp. causal) in the standard sense, $j = 1,\ldots,N$.
\item A piecewise smooth curve $\bgamma: I \rightarrow \MN$ is (future-directed) timelike (resp. causal) if its tangent vectors, whenever they exist, are (future-directed) timelike (resp. causal). This is equivalent to the requirement that the piecewise smooth curves $\bgamma^j \vc \iota^j \circ \bgamma$ are (future-directed) timelike (resp. causal) in the standard sense, $j = 1,\ldots,N$.
\item For any $\bp, \bq \in \MN$ we say that $\bp$ chronologically (resp. causally) precedes $\bq$ (symbolically $\bp \ll \bq$, resp. $\bp \preceq \bq$) if there exists a future-directed timelike (resp. causal) curve $\bgamma: [0,1] \rightarrow \MN$ such that $\bgamma(0) = \bp$ and $\bgamma(1) = \bq$. This is equivalent to the requirement that $\bp^j \ll \bq^j$ (resp. $\bp^j \preceq \bq^j$) in the standard sense, $j = 1,\ldots,N$. As before, we also write $\bp \prec \bq$ to denote $(\bp \preceq \bq \ \wedge \ \bp \neq \bq)$.
\item Extending (ii), we say that a continuous curve $\bgamma \in C( I, \MN )$ is future-directed causal if for any $s,t \in I$ the inequality $s < t$ implies that $\bgamma(s) \prec \bgamma(t)$. This is equivalent to the requirement that the curves $\bgamma^j \vc \iota^j \circ \bgamma \in C( I, \M )$ are future-directed causal in the standard sense, $j = 1,\ldots,N$.
\end{enumerate}

We shall use the symbols $I^+_{\mN}$ and $J^+_{\mN}$ to denote, respectively, the chronological and the causal precedence relations in the $N$-particle setting. In full analogy with their standard one-particle counterparts one might consider the future/past sets $I^\pm_{\mN}(\X), J^\pm_{\mN}(\X) \subset \MN$, as well as the sets $I^\pm_{\mN}(\bp, \U)$, where $\X, \U \subset \MN$. Observe that, by (iii) above,
\begin{align*}
\iota^j(I^\pm_{\mN}(\X)) = I^\pm(\iota^j(\X)) \qquad \textrm{and} \qquad \iota^j(J^\pm_{\mN}(\X)) = J^\pm(\iota^j(\X)).
\end{align*}
\end{definition}

For the sake of brevity, we shall from now on omit the term `future-directed' when referring to causal curves.

Some basic facts from the causality theory can be easily shown to hold also in the $N$-particle setting.
\begin{proposition}
\label{Ncausality_Prop1}
\
\begin{enumerate}[(i)]
\item Relation $I^+_{\mN}$ is irreflexive, transitive and open, whereas $J^+_{\mN}$ is a closed (a.k.a. continuous) partial order. Moreover, for any $\bp,\bq,\br \in \MN$ if $\bp \ll \bq \preceq \br$ or $\bp \preceq \bq \ll \br$, then $\bp \ll \br$.
\item For any $\bp \in \MN$ the set $I^\pm_{\mN}(\bp)$ is open with $\bp$ lying in its closure.
\item For any $\X \subset \MN$ the set $I^\pm_{\mN}(\X)$ is open.
\item For any compact $\K \subset \MN$ the set $J^\pm_{\mN}(\K)$ is closed.
\item For any compact $\K, \K' \subset \MN$ the set $J^+_{\mN}(\K) \cap J^-_{\mN}(\K')$ is compact.
\item For any compact $\K \subset \MN$ and $t \in \sR$ the set $J^\pm_{\mN}(\K) \cap \itSigma^N_t$ is compact.
\item For any compact $\K \subset \MN$ and $t \in \sR$ the set $J^\pm_{\mN}(\K) \cap J^\mp_{\mN}(\itSigma^N_t)$ is compact.
\end{enumerate}
\end{proposition}
\begin{pf}\textbf{.}
The order-theoretic properties listed in (i) hold on the strength of the very definition of $I^+_{\mN}$ and $J^+_{\mN}$ together with the fact that they hold for $I^+$ and $J^+$. As for the topological properties, observe that
\begin{align*}
I^+_{\mN} = \bigcap\limits_{j=1}^N (\iota^j, \iota^j)^{-1}(I^+) \quad \textnormal{and} \quad J^+_{\mN} = \bigcap\limits_{j=1}^N (\iota^j, \iota^j)^{-1}(J^+)
\end{align*}
and thus the openness of $I^+_{\mN}$ follows from the openness of $I^+$ and, similarly, the closedness of $J^+_{\mN}$ follows from the closedness of $J^+$, which in turn is guaranteed by the global hyperbolicity of $\M$.

To prove te first part of (ii), notice that, for any given $\bp$, also the set $I^\pm_{\mN}(\bp)$ can be expressed as a finite intersection of open sets, namely
\begin{align*}
I^\pm_{\mN}(\bp) = \bigcap\limits_{j=1}^N (\iota^j)^{-1}(I^\pm(\bp^j)).
\end{align*}
To see that $\bp \vc (t, x_1, \ldots, x_N)$ is in $\overline{I^\pm_{\mN}(\bp)}$, simply consider the sequence $((t \pm \tfrac{1}{k}, x_1, \ldots, x_N))_k$.

In order to obtain (iii), invoke (ii) and the fact that $I^\pm_{\mN}(\X) = \bigcup_{\bp \in \X} I^\pm_{\mN}(\bp)$.

To obtain (iv), observe first that $J^+_{\mN}(\K) = \pi^R( (\pi^L)^{-1}(\K) \cap J^+_{\mN} )$, where $\pi^L, \pi^R: \MN^2 \rightarrow \MN$ denote the canonical projections on the left and the right argument, respectively. Closedness follows from (i) and the fact that canonical projections are open maps. The reasoning for the causal pasts is analogous --- one may simply swap $\pi^L$ and $\pi^R$ above.

In order to prove (v), take any sequence $(\br_k) \subset J^+_{\mN}(\K) \cap J^-_{\mN}(\K')$. Since by (iv) the considered set is closed, our only task is to find its convergent subsequence. To this end, notice first that $(\br^j_k) \subset J^+(\iota^j(\K)) \cap J^-(\iota^j(\K'))$ for $j = 1,\ldots,N$. But the sets $J^+(\iota^j(\K)) \cap J^-(\iota^j(\K'))$ are all compact (see e.g. \cite[Lemma 11.5]{HR09}), therefore we can proceed as follows: First take a sequence $(k_l)$ of indices such that $(\br^1_{k_l})$ converges. Then, take a subsequence of the sequence $(k_l)$ such that the corresponding subsequence of $(\br^2_{k_l})$ converges as well. Repeating this procedure until $j=N$, we end up with a subsequence $(k_m)$ of indices such that the corresponding subsequences $(\br^j_{k_m})$ for all $j$'s converge. Invoking Proposition \ref{Ncausality_Prop0}, we obtain that the sequence $(\br_{k_m})$ is a convergent subsequence of $(\br_k)$, as desired. 

The proofs of (vi) and (vii) go along similar lines to the proof of (v) --- they also amount to finding convergent subsequences of elements from the considered sets. To see that in the case of (vi), observe that the sets $J^\pm_{\mN}(\K) \cap \itSigma^N_t$ are closed and that the sets $\iota^j(J^\pm_{\mN}(\K) \cap \itSigma^N_t) = J^\pm(\iota^j(\K)) \cap \T^{-1}(t)$ are compact for every $j=1,\ldots,N$ (see e.g. \cite[Property 4, p.44]{MS08}). Analogously, in the case of (vii), observe that
\begin{align*}
& J^+_{\mN}(\K) \cap J^-_{\mN}(\itSigma^N_t) = J^+_{\mN}(\K) \cap ((-\infty,t] \times \itSigma^N)
\\
\textrm{and} \quad & J^-_{\mN}(\K) \cap J^+_{\mN}(\itSigma^N_t) = J^-_{\mN}(\K) \cap ([t,+\infty) \times \itSigma^N),
\end{align*} 
hence the sets $J^\pm_{\mN}(\K) \cap J^\mp_{\mN}(\itSigma^N_t)$ are also closed, whereas the sets
\begin{align*}
\iota^j(J^\pm_{\mN}(\K) \cap J^\mp_{\mN}(\itSigma^N_t)) = J^\pm(\iota^j(\K)) \cap J^\mp(\T^{-1}(t))
\end{align*}
are compact for every $j=1,\ldots,N$ (see \cite[Proposition 13 (iii)]{Miller17a}).
\end{pf}

Let us remark that fact (i) states, in particular, that $\MN$ is a causal space in the sense of Kronheimer and Penrose \cite{KP67}.

Proposition \ref{Ncausality_Prop1} shows that the $N$-particle spacetime $\MN$ inherits, in a fairly straighforward manner, the standard properties of the causal structure of $\M$. It is less apparent that $\MN$ admits a sound notion of a ($N$-particle) Cauchy hypersurface. We present it below, following (and accordingly modifying) the exposition by O'Neill \cite{BN83}, and show that it exhibits the usual topological and causal properties. Before we embark on this quest, let us make the following useful observation.
\begin{proposition}
\label{Ncausality_Prop2}
Any inextendible causal (in particular, timelike) curve $\bgamma: (a,b) \rightarrow \MN$, where $-\infty \leq a < b \leq +\infty$, can be reparametrised by the time parameter associated with the chosen GBS splitting, i.e., there exists a continuous strictly increasing map $\rho: \sR \rightarrow (a,b)$ such that $\pi^0 \circ \bgamma \circ \rho = \textnormal{id}_\sR$. In other words, any such a curve can be regarded as a map $\sR \ni s \mapsto (s, x_1(s), \ldots, x_N(s)) \in \MN$.
\end{proposition}
\begin{pf}\textbf{.}
Let $\bgamma: (a,b) \rightarrow \MN$ be an inextendible causal curve. This implies that $\bgamma^j$ is also an inextendible causal curve in $\M$ for every $j=1,\ldots,N$. But since $\pi^0 = \T \circ \iota^j$, we have that $\pi^0 \circ \bgamma = \T \circ \bgamma^j$ (for every $j$). Because $\T$ is a Cauchy temporal function, this implies that $\pi^0 \circ \bgamma$ is actually a continuous strictly increasing map onto $\sR$, and so $(\pi^0 \circ \bgamma)^{-1}: \sR \rightarrow (a,b)$ is a well-defined reparametrisation map with the desired property.
\end{pf}

\begin{definition}
\label{Cauchy_Def}
A subset $\cS \subset \MN$ is called a \emph{Cauchy hypersurface} if it is met exactly once by every inextendible $N$-particle timelike curve.
\end{definition}
Obviously, $\MN$ admits Cauchy hypersurfaces --- the subsets $\itSigma^N_t$ for any $t \in \sR$ provide immediate examples. However, there exist many more of them and we would like to know how similar they work compared to their standard counterparts.
\begin{proposition}
\label{Ncausality_Prop3}
Any Cauchy hypersurface $\cS \subset \MN$ is a closed achronal topological hypersurface. Furthermore, $\cS$ is connected.
\end{proposition}
\begin{pf}\textbf{.}
Achronality follows trivially from the definition. The rest of the proof closely follows O'Neill \cite[pp. 413--417]{BN83}.

To prove that $\cS$ is closed, take any $\bp \in \MN$ and consider an inextendible timelike curve passing through it. The fact that this curve meets $\cS$ exactly ones means that $\bp \in I^-_{\mN}(\cS) \cup \cS \cup I^+_{\mN}(\cS)$ with the three subsets being disjoint. Hence $\cS = \MN \setminus ( I^-_{\mN}(\cS) \cup I^+_{\mN}(\cS) )$ is closed, being a complement of an open set.

To prove that $\cS$ is a topological hypersurface, we need to show that for any $\bp \vc (t,x_1,\ldots,x_N)$ $\in \cS$ there exists a neighborhood $\Uf$ of $\bp$ in $\MN$ and a homeomorphism $\phi$ mapping $\Uf$ into an open subset of $\sR^{N(n-1)+1}$ such that $\phi(\Uf \cap \cS)$ is contained in a hyperplane. 

To this end, let $(a_j-\delta_j, b_j+\delta_j) \times V_j \subset \sR \times \itSigma$ be a coordinate neighborhood of $(t,x_j)$ such that $\{a_j\} \times V_j \subset I^-((t,x_j), \sR \times V_j)$ and $\{b_j\} \times V_j \subset I^+((t,x_j), \sR \times V_j)$, for each $j = 1,\ldots,N$ (we omit writing $\Phi^{-1}$ for brevity). Taking $a \vc \min_j \{a_j\}$, $b \vc \max_j \{b_j\}$ and $\delta \vc \max_j \{\delta_j\}$ we have that the set $\Uf \vc (a-\delta, b+\delta) \times \prod_{j=1}^N V_j$ is a neighborhood of $\bp$ with the property that
\begin{align}
\label{Ncausality_Prop3a}
\{a\} \times \prod_{j=1}^N V_j \subset I^-_{\mN}\left(\bp, \, \sR \times \prod_{j=1}^N V_j\right) \ \textnormal{and} \ \, \{b\} \times \prod_{j=1}^N V_j \subset I^+_{\mN}\left(\bp, \, \sR \times \prod_{j=1}^N V_j\right)
\end{align}
Define the map $h: \prod_{j=1}^N V_j \rightarrow (a,b)$ as follows: $h(y_1,\ldots,y_N)$ is the time coorditate of the unique intersection point of $\cS$ and the timelike curve $s \mapsto (s,y_1,\ldots,y_N)$. Notice that $h$ is well defined by (\ref{Ncausality_Prop3a}). We now assert that the map $\phi: \Uf \rightarrow \sR^{N(n-1)+1}$ defined via
\begin{align*}
\phi(s,y_1,\ldots,y_N) \vc \left( s - h(y_1,\ldots,y_N), \xi^1(y_1), \ldots, \xi^N(y_N) \right),
\end{align*}
with $\xi^j: V_j \rightarrow \sR^{n-1}$ being charts on $V_j \subset \itSigma$ for $j=1,\ldots,N$, is the desired homeomorphism which maps $\Uf \cap \cS$ into the hyperplane $\{0\} \times \sR^{N(n-1)}$. Clearly, $\phi$ is invertible and in order to prove the continuity of $\phi$ and $\phi^{-1}$ it suffices to show that $h$ is continuous.

Suppose it is not, i.e., there exists a sequence $(y_{1,k},\ldots, y_{N,k})_k$ convergent to some $(y_1,\ldots,y_N)$ $\in \prod_{j=1}^N V_j$ such that $h(y_{1,k},\ldots, y_{N,k}) \not\rightarrow h(y_1,\ldots,y_N)$. Since the latter sequence of real numbers is bounded (by $a$ from below and by $b$ from above), it has a subsequence convergent to some $h_0 \in [a,b]$. But then, by the very definition of $h$, $(h_0,y_1,\ldots,y_N)$ is either in the chronological future or the chronological past of $\bq \vc \left( h(y_1,\ldots,y_N), y_1,\ldots,y_N \right) \in \cS$:
\begin{align*}
\left( h_0, y_1,\ldots,y_N \right) \in I^+_{\mN}\left( \bq, \Uf \right) \cup I^-_{\mN}\left( \bq, \Uf \right).
\end{align*}
However, the above union is open in $\Uf$ and thus also
\begin{align*}
\left(h(y_{1,k},\ldots, y_{N,k}), y_{1,k},\ldots,y_{N,k}\right) \in I^+_{\mN}\left( \bq, \Uf \right) \cup I^-_{\mN}\left( \bq, \Uf \right).
\end{align*}
for all but finitely many $k$. But this blatantly contradicts the achronality of $\cS$. This concludes the proof of the continuity of $h$ together with the proof of $\cS$ being a topological hypersurface.

Finally, to prove that $\cS$ is connected, let us define the map $\psi: \cS \times \sR \rightarrow \MN$ via $\psi((t,x_1,\ldots,x_N),s) \vc (t+s, x_1,\ldots,x_N)$. Clearly, $\psi$ is continuous, and the assumption that $\cS$ is a Cauchy hypersurface guarantees it is also one-to-one and onto. Since $\cS \times \sR$ and $\MN$ are topological manifolds of the same dimension (as attested by the earlier part of the proof), by the invariance of domain we obtain that $\psi$ is a homeomorphism.

Define $\theta \vc \pi_{\cS} \circ \psi^{-1}: \MN \rightarrow \cS$ with $\pi_{\cS}: \cS \times \sR \rightarrow \cS$ denoting the canonical projection onto the first argument. The map $\theta$ is an open, continuous surjection, and since $\MN$ is connected (being a product of connected spaces), then so is $\cS$.
\end{pf}

Observe, additionally, that the map $\theta$ just defined is a retraction, i.e., that $\theta|_{\cS} = \textnormal{id}_\cS$, what allows us to obtain a straightforward analogue of \cite[Corollary 14.32]{BN83}.
\begin{corollary}
Any two Cauchy hypersurfaces $\cS$, $\cS'$ in $\MN$ are homeomorphic.
\end{corollary}
\begin{pf}\textbf{.}
Let $\theta$, $\theta'$ be the retractions onto $\cS$ and $\cS'$, respectively, as defined above. Then $\theta|_{\cS'}$ and $\theta'|_{\cS}$ can be easily shown to be mutually inverse maps.
\end{pf}

To finish this subsection, let us notice that also the notions of causal and (Cauchy) time functions naturally generalise to the $N$-particle setting. Concretely, a function $\tau: \MN \rightarrow \sR$ is referred to as
\begin{itemize}
\item a \emph{causal} function if $\bp \preceq \bq$ implies $\tau(\bp) \leq \tau(\bq)$,
\item a \emph{time} function if it is continuous and $\bp \prec \bq$ implies $\tau(\bp) < \tau(\bq)$,
\item a \emph{Cauchy} time function if it is a time function whose level sets are all Cauchy hypersurfaces in $\MN$.
\end{itemize}
Notice, on the other hand, that the notion of a temporal function is not immediately generalisable, because on $\MN$ there is no natural gradient (recall that we have not equipped $\MN$ with the Lorentzian structure).


\section{\texorpdfstring{Causality between $N$-particle measures}{Causality between N-particle measures}}\label{sec:NJplus}

\subsection{Causal precedence via causal couplings}

Let $\Pf(\MN)$ denote the space of all Borel probability measures on the `$N$-particle configuration spacetime' $\MN$, which we shall be calling ``$N$-particle measures'' or simply ``measures'' from now on. Any such a measure $\bmu$ encodes not only the bare probability densities of each of the $N$ particles, but also the correlations between them. For instance, the product measure
\begin{align}\label{disting}
\bmu = \delta_t \x \mu_1 \x \mu_2, \quad \text{ with } \quad \mu_1, \mu_2 \in \Pf(\itSigma)
\end{align}
models the probability distribution of $N=2$ uncorrelated distinguishable particles at time instant $t$. On the other hand, a symmetric measure
\begin{align}\label{indisting}
\bmu = \delta_t \x \tfrac{1}{2} \left( \mu_1 \x \mu_2 + \mu_2 \x \mu_1 \right), \quad \text{ with } \quad \mu_1, \mu_2 \in \Pf(\itSigma)
\end{align}
describes a pair of two indistinguishable entities. Obviously, the space $\Pf(\MN)$ contains many more general elements, which can be thought of as modelling `partially distinguishable particles'.

Drawing from the optimal transport theory, it is possible to extend the causal precedence relation introduced on $\MN$ onto $\Pf(\MN)$. The $N = 1$ case has been put forward and extensively studied in \cite{AHP2017} and, indepedently, in \cite{Suhr2016}.
\begin{definition}\label{def:caus}
For any $\bmu, \bnu \in \Pf(\MN)$, we say that $\bmu$ \emph{causally precedes} $\bnu$ (denoted $\bmu \preceq \bnu$) if there exists $\bomega \in \Pf(\MN^2)$ such that
\begin{enumerate}[i)]
\item $\pi^L_\sharp \bomega = \bmu$ and $\pi^R_\sharp \bomega = \bnu$,
\item $\bomega(J^+_{\mN}) = 1$,
\end{enumerate}
where $\pi^L, \pi^R: \MN^2 \rightarrow \MN$ denote the canonical projections on the left and the right argument, respectively.
\end{definition}

In the language of optimal transport, any $\bomega \in \Pf(\MN^2)$ fulfilling condition (i) is called a \emph{coupling} of $\bmu$ and $\bnu$, which in turn are $\bomega$'s left and right \emph{marginals}. If condition (ii) is fulfilled as well, we call such an $\bomega$ a \emph{causal coupling}. For future use, denote the set of all couplings (resp. causal couplings) of $\bmu$ and $\bnu$ by $\itPi(\bmu,\bnu)$ (resp. $\itPi_\preceq(\bmu,\bnu)$).

Couplings, also known as \emph{tranference plans}, have been originally introduced by Kantorovich in his seminal treatment of the optimal transportation problem \cite{Kantorovich1,Kantorovich2} (see e.g. \cite{CV03} for an excellent introduction to this topic). Intuitively speaking, a coupling of the probability measures $\bmu$ and $\bnu$ describes how to ``reconfigure'' the former into the latter. This ``reconfiguration'' involves transporting the (possibly infinitesimal) ``portions of probability'' between points of $\MN$, and the coupling $\bomega \in \itPi(\bmu,\bnu) \subset \Pf(\MN^2)$ tells precisely what amount of probability is transported between any given pair of points. Crucially, if the coupling is causal, then the transportation of probability is allowed only between causally related pairs of points or, equivalently, 
\begin{quote}
\textit{Each infinitesimal portion of probability must travel along a causal curve.}
\end{quote}
Let us emphasise that the current $N$-particle setting contains an important conceptual novelty, which was absent in the single-particle case. Namely, the above ``hydrodynamic'' interpretation concerns not just the causal flow of individual particles' probability densities, but also of the \emph{correlations} in the system.

Definition \ref{def:caus} extends the causal precedence relation between the points of $\MN$ in the following sense: For any $\bp,\bq \in \MN$ the Dirac measures $\delta_{\bp}, \delta_{\bq}$ satisfy $\delta_{\bp} \preceq \delta_{\bp}$ iff $\bp \preceq \bq$ in the sense of Definition \ref{Ncausality_Def} (iii). Indeed, the only coupling of $\delta_{\bp}$ and $\delta_{\bq}$ is $\bomega := \delta_{\bp} \times \delta_{\bq} = \delta_{(\bp,\bq)}$ \cite[Proposition 4]{AHP2017}, and this coupling is causal iff $\bp \preceq \bq$ (cf. \cite[Corollary 5]{AHP2017}).

The extended relation $\preceq$ is a partial order on $\Pf(\MN)$, just like in the single-particle case \cite{AHP2017}.
\begin{proposition}
The relation $\preceq$ is reflexive, transitive and antisymmetric.
\end{proposition}
\begin{pf}\textbf{.}
The adaptation of the proofs of \cite[Theorems 11 \& 12]{AHP2017} is straightforward. The only point requiring a comment is how to define, in the present context, the function $f_\K: \MN \rightarrow \sR$ for a given compact subset $\K \subset \MN$, appearing in the proof of \cite[Theorem 12]{AHP2017}. One possibility is simply
\begin{align*}
    f_\K(t,x_1,\ldots,x_N) \vc \begin{cases}
                            \arctan t & \textnormal{for } (t,x_1,\ldots,x_N) \in \K \\
                            0 & \textnormal{for } (t,x_1,\ldots,x_N) \not\in \K
                            \end{cases}.
\end{align*}
Indeed, thus defined $f_\K$ is Borel, bounded, and has the property that $f_\K(\bp) < f_\K(\bq)$ provided $\bp,\bq \in \K$ and $\bp \prec \bq$, just as needed in the above-mentioned proof in \cite{AHP2017}.
\end{pf}


\subsection{Equivalent characterisations}\label{sec::equiv}

Definition \ref{def:caus} admits a number of equivalent formulations, in close analogy with the single-particle case (cf. \cite[Theorems 8 \& 10]{AHP2017}). They all express the causal relation between measures in terms of inequalities between real numbers.

\begin{theorem}\label{thm::equiv} For any $\bmu, \bnu \in \Pf(\MN)$ the following conditions are equivalent:
\begin{enumerate}[1{$^\circ$}]
\item $\bmu \preceq \bnu$
\item For any compact subset $\K \subset \MN$
    \begin{align}
    \label{causal2}
    \bmu(J^+_{\mN}(\K)) \leq \bnu(J^+_{\mN}(\K))
    \end{align}
\item For any Borel subset $\F \subset \MN$ satisfying $J^+_{\mN}(\F) \subset \F$
    \begin{align}
    \label{causal3}
    \bmu(\F) \leq \bnu(\F)
    \end{align}
\item For any compact subset $\K \subset \supp \bmu$
    \begin{align}
    \label{causal4}
    \bmu(\K) \leq \bnu(J^+_{\mN}(\K)).
    \end{align}
\item For any Cauchy hypersurface $\cS \subset \MN$
    \begin{align}
    \label{causal5}
    \bmu(J^+_{\mN}(\cS)) \leq \bnu(J^+_{\mN}(\cS))
    \end{align}
\item For any bounded time function $\tau$
    \begin{align}
    \label{causal6}
    \int_{\MN} \tau d\bmu \leq \int_{\MN} \tau d\bnu.
    \end{align}
\end{enumerate}
\end{theorem}
Conditions {\it 2{$^\circ$}}, {\it 3{$^\circ$}}, {\it 5{$^\circ$}} reduce the transport-theoretic meaning of ``$\bmu \preceq \bnu$'' to numerical relations between the $\bmu$-measures and $\bnu$-measures of certain (classes of) future sets. For example, condition {\it 5{$^\circ$}} says, loosely speaking, that there is ``more of $\bnu$ than of $\bmu$'' in the future of any $N$-particle Cauchy hypersurface.

Condition {\it 4{$^\circ$}} possesses a particularly lucid physical interpretation, closely related to the ``hydrodynamic'' picture sketched above. Namely, it says that the probability ``mass'' contained within any given compact subset of space cannot escape the latter's causal future. It first appeared in \cite{Suhr2016} and, independently, in \cite{PRA2017}. However, similar approach to causality was employed already in \cite{Gerlach1969,Gerlach1968,Gromes1970}, where the concept of ``causal propagation of observables'' was put forward, as well as in the works of Hegerfeldt \cite{Hegerfeldt1,Hegerfeldt1985,HegerfeldtFermi,Hegerfeldt2}. Crucially, condition {\it 4{$^\circ$}} allows for an operational interpretation in terms of local detection statistics \cite{PRA2020}.

Finally, condition {\it 6{$^\circ$}} connects with the ``dual approach'' to recover the causal order from a specific set of functions \cite{Bes09,CQG2013,MinguzziUtilities}. The latter is a starting point for the study of causality in noncommutative spacetimes from a $C^*$-algebraic perspective \cite{Bes09,BesReview,UNIV2017,CQG2013,Moretti}.

\begin{pf}\textbf{ of Theorem \ref{thm::equiv}.}
Since $J^+_{\mN}$ is closed (Proposition \ref{Ncausality_Prop1} (i)), equivalences $1^\circ \Leftrightarrow 2^\circ \Leftrightarrow 3^\circ$ follow from \cite[Theorem 4]{Miller18}, where they were proven in the broader context of closed preorders. The implication $2^\circ \Rightarrow 4^\circ$ follows from the obvious inequality $\bmu(\K) \leq \bmu(J^+_{\mN}(\K))$. We also have, trivially, that $3^\circ \Rightarrow 5^\circ$.

To prove $4^\circ \Rightarrow 2^\circ$, let $\K$ be any compact subset of $\MN$ and recall that, by Ulam's tightness theorem, the $\bmu$-measure of any Borel set $\U \subset \MN$ can be approximated from below by the $\bmu$-measures of its compact subsets. In particular, putting $\U \vc J^+_{\mN}(\K) \cap \supp \bmu$, one has that for any $\varepsilon > 0$ there exists a compact set $\mathcal{C}_\varepsilon \subset J^+_{\mN}(\K) \cap \supp \bmu$ such that $\bmu(J^+_{\mN}(\K) \cap \supp \bmu) \leq \bmu(\mathcal{C}_\varepsilon) + \varepsilon$.

Using $4^\circ$, one thus can write that
\begin{align*}
\bmu(J^+_{\mN}(\K)) & = \bmu(J^+_{\mN}(\K) \cap \supp \bmu) \leq \bmu(\mathcal{C}_\varepsilon) + \varepsilon \leq \bnu(J^+_{\mN}(\mathcal{C}_\varepsilon)) + \varepsilon 
\\
& \leq \bnu(J^+_{\mN}(J^+_{\mN}(\K))) + \varepsilon = \bnu(J^+_{\mN}(\K)) + \varepsilon,
\end{align*}
where in the last equality we used the transitivity of $\preceq$. Taking now $\varepsilon \rightarrow 0^+$ yields $2^{\circ}$.

We now move to proving $5^\circ \Rightarrow 2^\circ$, to which end we adapt the proof of Theorem 10 in \cite{AHP2017}.

Let $\K \subset \MN$ be any compact set and let $t_0 \vc \min_{\bp \in \K} \pi^0(\bp)$. For every $k \in \sN$ define
\begin{align*}
\cS_k \vc \partial J^+_{\mN}(\K \cup \itSigma^N_{t_0+k}) = \partial \left[ J^+_{\mN}(\K) \cup ([t_0+k,\infty) \times \itSigma^N)\right].
\end{align*}
We assert that $\cS_k$ is a Cauchy hypersurface and that
\begin{align}
\label{equiv_proof1}
J^+_{\mN}(\cS_k) = J^+_{\mN}(\K \cup \itSigma^N_{t_0+k}) = J^+_{\mN}(\K) \cup ([t_0+k,\infty) \times \itSigma^N).
\end{align}
First, observe that every inextendible timelike curve $\bgamma$ meets $\cS_k$. Indeed, if we parametrise the curve with the time coordinate $t$, then
\begin{align*}
& \bgamma(t) \not\in J^+_{\mN}(\K) \cup ([t_0+k,\infty) \times \itSigma^N) \quad \textrm{for} \ t < t_0 \quad \textrm{and}
\\
& \bgamma(t) \in J^+_{\mN}(\K) \cup ([t_0+k,\infty) \times \itSigma^N) \quad \textrm{for} \ t \geq t_0 + k,
\end{align*}
therefore $\bgamma$ must cross the boundary.

In order to demonstrate that $\cS_k$ is a Cauchy hypersurface, it now suffices to prove its achronality, to which end we mimic the first part of the proof of \cite[Chapter 14, Corrolary 27]{BN83}. For convenience, denote $\J \vc J^+_{\mN}(\K \cup \itSigma^N_{t_0+k})$ and let $\bp \in \partial \J$. If $\bq \in I^+_{\mN}(\bp)$, then $I^-_{\mN}(\bq)$ is an open neighborhood of $\bp$ and hence $I^-_{\mN}(\bq) \cap \J \neq \emptyset$, which means that $\bq \in I^+_{\mN}(\J) \subset \J$. We have thus obtained that $I^+_{\mN}(\bp) \subset \J$. Dually, one can show that $I^-_{\mN}(\bp) \subset \J^c$. Altogether, we get that $I^+_{\mN}(\partial \J) \cap I^-_{\mN}(\partial \J) \subset \J \cap \J^c = \emptyset$ and so $\partial \J = \cS_k$ is achronal.

To show \eqref{equiv_proof1}, we invoke \cite[Lemma 2]{AHP2017} adapted to the $N$-particle setting: If $\F \subset \MN$ is a closed set such that $J^+_{\mN}(\F) \subset \F \subset J^+_{\mN}(\X)$ for some achronal set $\X \subset \MN$, then $J^+_{\mN}(\partial \F) = \F$. The proof provided in \cite{AHP2017} remains valid in the present setting (in particular thanks to Proposition \ref{Ncausality_Prop1} (i)).

The set $\J$ clearly satisfies the assumptions of the above-mentioned lemma --- it is closed and it satisfies $J^+_{\mN}(\J) \subset \J \subset J^+_{\mN}(\itSigma^N_{t_0+k})$. On the strength of the lemma we thus obtain \eqref{equiv_proof1}.

By $5^\circ$ we have that $\bmu(J^+_{\mN}(\cS_k)) \leq \bnu(J^+_{\mN}(\cS_k))$ for all $k \in \sN$ and since $J^+_{\mN}(\cS_k) \supset J^+_{\mN}(\cS_{k+1})$, we can write
\begin{align*}
\bmu\left( \bigcap\limits_{k=0}^\infty J^+_{\mN}(\cS_k)\right) \leq \bnu\left( \bigcap\limits_{k=0}^\infty J^+_{\mN}(\cS_k)\right).
\end{align*}
It now remains to realise that the intersection is nothing but $J^+_{\mN}(\K)$. Indeed,
\begin{align*}
\bigcap\limits_{k=0}^\infty J^+_{\mN}(\cS_k) = J^+_{\mN}(\K) \cup \underbrace{\bigcap\limits_{k=0}^\infty [t_0+k,\infty) \times \itSigma^N}_{=\, \emptyset} = J^+_{\mN}(\K).
\end{align*}

As far as $6^\circ$ is concerned, let us first notice that $1^\circ$ implies it almost immediately --- simply take any $\bomega$ as specified by Definition \ref{def:caus} and write
\begin{align*}
\int_{\MN} \tau d\bmu = \int_{J^+_{\mN}} \tau(\bp) d\bomega(\bp,\bq) \leq \int_{J^+_{\mN}} \tau(\bq) d\bomega(\bp,\bq) = \int_{\MN} \tau d\bnu,
\end{align*}
where we have only used the fact that $\tau$ is a causal function. 

Conversely, in order to show that $6^\circ \Rightarrow 5^\circ$, fix a Cauchy hypersurface $\cS$ and for any $k \in \sN$ define
\begin{align*}
\tau_k \vc \varphi_k \circ \pi_\sR \circ \psi^{-1},
\end{align*}
where $\psi: \cS \times \sR \rightarrow \MN$ is the homeomorphism that featured in the last part of the proof of Proposition \ref{Ncausality_Prop3}, $\pi_\sR: \cS \times \sR \rightarrow \sR$ is the canonical projection, and $\varphi_k \in C^\infty(\sR)$ is given by the formula $\varphi_k(x) = \tfrac{1}{2} + \tfrac{1}{2} \tanh(k^2x+k)$. Observe that all $\varphi_k$'s are positive, strictly increasing and bounded by $1$, and that the sequence $(\varphi_k)$ converges pointwise to the characteristic function of $[0,\infty)$.

One can easily convince oneself that every $\tau_k$ is a positive time function bounded by $1$, and so by $6^\circ$
\begin{align*}
\forall k \in \sN \qquad \int_{\MN} \tau_k d\bmu \leq \int_{\MN} \tau_k d\bnu.
\end{align*}
Invoking Lebesgue's dominated convergence theorem and noticing that $\pi_\sR(\psi^{-1}(\bp)) \geq 0$ iff $\bp \in J^+_{\mN}(\cS)$, in the limit $k \rightarrow \infty$ one obtains the desired inequality \eqref{causal5}.
\end{pf}

\begin{remark}\label{rem6}
Without any changes in the proof, in condition $6^\circ$ the word ``bounded'' could be replaced with ``smooth bounded'' or ``$\bmu$- and $\bnu$- integrable'', whereas ``time'' could be weakened to ``continuous causal'' or just ``causal'' (cf. \cite{AHP2017,Miller17}).  
\end{remark}


\section{Causal evolution of measures}\label{sec:evo}

Let us now turn to the question of a causal time-evolution of measures. In order to guide the intuition and introduce the necessary notions we briefly recall the main results of \cite{Miller17a} concerning the single-particle case.

Recall first that $\T$ is a fixed Cauchy temporal function determining the GBS splitting $\Phi: \M \rightarrow \sR \times \itSigma$ such that $\Phi(\T^{-1}(t)) = \{t\} \times \itSigma \cv \itSigma_t$. Let also $I$ be a fixed interval. 

By an \emph{evolution of measures} we understand any measure-valued map $I \ni t \mapsto \mu_t \in \Pf(\M)$ such that $\supp \mu_t \subset \T^{-1}(t)$. We say that the evolution is \emph{causal} if
\begin{align}
\label{causal_evo}
\forall \, s,t \in I \quad s \leq t \ \Rightarrow \ \mu_s \preceq \mu_t
\end{align} 
in the sense of Definition \ref{def:caus} (for $N=1$).

The evolution of measures thus defined seems observer-dependent, as it explicitly refers to the Cauchy temporal function $\T$ and the GBS splitting induced by it. It is by no means clear how two observers $O$ and $O'$, employing two different GBS splittings induced by $\T$ and $\T'$, respectively, would be able to compare their respective families of measures $\{\mu_t\}$ and $\{\mu'_\tau\}$. In \cite{Miller17a} it was proven, however, that the causal evolution of measures can be in fact described in an equivalent, yet manifestly observer-independent manner, which involves a suitably topologised space of causal curves. We briefly discuss the question of splitting-independence in the end of the next subsection.

By $C^I_\T$ we denote the space of all continuous future-directed causal curves $\gamma: I \rightarrow \M$ such that $\exists c_\gamma > 0$ $\forall s,t \in I \quad \T(\gamma(t)) - \T(\gamma(s)) = c_\gamma (t-s)$, endowed with the compact-open topology induced from $C(I,\M)$.

One can show that thus defined $C^I_\T$ is a Polish space \cite[Proposition 4]{Miller17a}. Even more importantly, one has the following equivalence.
\newpage
\begin{theorem}[\!\!\cite{Miller17a}]
\label{PolishThm}
Let $I \ni t \mapsto \mu_t$ be an evolution of measures. The following conditions are equivalent:
\begin{enumerate}[i)]
\item The evolution $t \mapsto \mu_t$ is causal (i.e. it satisfies \eqref{causal_evo}).
\item There exists $\sigma \in \Pf(C^I_\T)$ such that $(\ev_t)_\sharp \sigma = \mu_t$ for every $t \in I$, where \\$\ev_t: C^I_\T \rightarrow \M$ denotes the evaluation map.
\end{enumerate}
\end{theorem}

Theorem \ref{PolishThm} thus allows to reexpress any given causal evolution of measures $\mu_t$ as a single measure $\sigma$ on the space of causal curves, from which $\mu_t$'s can be retrieved when needed. Moreover, because causal curves (after ``deparametrisation'') are GBS-splitting-in\-de\-pen\-dent objects, one can thus regard $\sigma$ as providing an observer-in\-de\-pen\-dent description of a time-evolving physical entity. To put it differently: Just as a single causal curve is a geometrical object describing the time-evolution of a pointlike particle, a measure on the space of causal curves is a geometrical object describing the time-evolution of a measure-like entity. For a more detailed discussion the Reader is referred to \cite[Section 2]{Miller17a}.


\subsection{\texorpdfstring{Spaces of $N$-particle causal curves}{Spaces of N-particle causal curves}}

In order to move forward into the $N$-particle setting we need to identify the suitable spaces of $N$-particle causal curves. To this end, let us introduce two additional closely related auxiliary spaces.

Firstly, let $A^I_\T$ be the subspace of $C^I_\T$ containing those causal curves $\gamma$ which satisfy $\T \circ \gamma = \id_I$. The set $A^I_\T$ is a closed subspace of the Polish space $C^I_\T$ and hence it is itself Polish.

Secondly, define $B^I_\T \vc \{ r \in C(I, \itSigma) \; | \; \textrm{the curve } t \mapsto \Phi^{-1}(t,r(t)) \textrm{ is causal} \}$, endowed with the compact-open topology induced from $C(I, \itSigma)$.

In order to better grasp the topologies of $A^I_\T$ and $B^I_\T$, let us fix a distance function $d_\itSigma$ on $\itSigma$ and with its help define a distance function $d$ on $\sR \times \itSigma$ via $d((s,x),(t,y))\vc |s - t| + d_\itSigma(x,y)$. Then, as a distance function on $\M$ let us take the pullback $\Phi^\ast d$. With these distance functions at hand, convergent sequences in $A^I_\T$ and $B^I_\T$ are exactly those which converge uniformly on compact (sub)intervals:
\begin{align*}
\gamma_k \rightarrow \gamma \textrm{ in } A^I_\T \qquad & \Leftrightarrow \qquad \forall [a,b] \subset I \quad \sup_{t \in [a,b]} \Phi^\ast d(\gamma_k(t), \gamma(t)) \rightarrow 0,
\\
r_k \rightarrow r \textrm{ in } B^I_\T \qquad & \Leftrightarrow \qquad \forall [a,b] \subset I \quad \sup_{t \in [a,b]} d_\itSigma(r_k(t), r(t)) \rightarrow 0.
\end{align*}

\begin{proposition}
\label{NcurvesAB}
The spaces $A^I_\T$ and $B^I_\T$ are homeomorphic.
\end{proposition}
\begin{pf}\textbf{.}
Define the maps $F: A_\T^I \rightarrow B_\T^I$ and $G: B_\T^I \rightarrow A_\T^I$ via
\begin{align*}
F(\gamma) \vc \pi_\itSigma \circ \Phi \circ \gamma \quad \textrm{and} \quad G(r) \vc \Phi^{-1} \circ (\id_I,r),
\end{align*}
where $\pi_\itSigma: \sR \times \itSigma \rightarrow \itSigma$ is the canonical projection. It is easy to check that $F$ and $G$ are well defined and mutually inverse. In order to prove they are continuous, observe that
\begin{align*}
\Phi^\ast d(G(r_1)(t), G(r_2)(t)) = d((t,r_1(t)), (t,r_2(t))) = d_\itSigma(r_1(t), r_2(t))
\end{align*} 
for any $r_1,r_2 \in B_\T^I$ and any $t \in I$. Hence, $G$ maps convergent sequences in $B_\T^I$ to convergent sequences in $A_\T^I$. Of course, its inverse $F$ enjoys analogous property. 
\end{pf}

\begin{definition}
\label{NcurvesDef}
Let $\Gamma^I_{\mN}$ be the space of all continuous causal curves $\bgamma: I \rightarrow \MN$ such that $\pi^0 \circ \bgamma = \id_I$, endowed with the compact-open topology induced from $C(I,\MN)$.
\end{definition}

Notice that every $\bgamma \in \Gamma^I_{\mN}$ has the form $\bgamma(t) = (t, r_1(t), \ldots, r_N(t))$ with each $r_j \vc F(\iota^j \circ \bgamma) = F(\bgamma^j)$ belonging to $B_\T^I$. Thus the space $\Gamma^I_{\mN}$ might be defined more succinctly as $\{\id_I\} \times (B_\T^I)^N$. The convergence of sequences in $\Gamma^I_{\mN}$ can be characterised via
\begin{align}
\label{Nconvergence}
\bgamma_k \rightarrow \bgamma \quad \textrm{in } \Gamma^I_\T & \quad \Leftrightarrow \quad \bgamma_k^j \rightarrow \bgamma^j \quad \ \ \textrm{in } A^I_\T \textrm{ for all } j = 1,\ldots,N
\\
\nonumber
& \quad \Leftrightarrow \quad (r_k)_j \rightarrow r_j \quad \textrm{in } B^I_\T \textrm{ for all } j = 1,\ldots,N.
\end{align}
The above characterisation of convergence can be viewed as an analogue of Proposition \ref{Ncausality_Prop0} for $N$-particle causal curves.

\begin{proposition}
\label{NcurvesProp1}
$\Gamma^I_{\mN}$ is a Polish space.
\end{proposition}
\begin{pf}\textbf{.}
On the strength of \eqref{Nconvergence} and Proposition \ref{NcurvesAB}, one has that $\Gamma^I_{\mN} \cong (B_\T^I)^N \cong (A^I_\T)^N$. The claim follows from the Polishness of $A^I_\T$.
\end{pf}

For the next result, recall from \cite[Proposition 3]{Miller17a} that for any compact $K_1,K_2 \subset \M$ the set $C^{[a,b]}_\T(K_1,K_2) \vc C^{[a,b]}_\T \cap \ev_a^{-1}(K_1) \cap \ev_b^{-1}(K_2)$ is compact. The same is true for the set $A^{[a,b]}_\T(K_1,K_2) \vc C^{[a,b]}_\T(K_1,K_2) \cap A^{[a,b]}_\T$, being a closed subset of a compact space.
\begin{proposition}
\label{NcurvesProp3}
Let $\K_1, \K_2 \subset \MN$ be compact and let $\Gamma^{[a,b]}_{\mN}(\K_1, \K_2)$ be the set of all $\bgamma \in \Gamma^{[a,b]}_{\mN}$ such that $\bgamma(a) \in \K_1$ and $\bgamma(b) \in \K_2$. Then $\Gamma^{[a,b]}_{\mN}(\K_1, \K_2)$ is compact.
\end{proposition}
\begin{pf}\textbf{.}
By the continuity of the evaluation maps, the set $\Gamma^{[a,b]}_{\mN}(\K_1, \K_2)$ is closed, and so our only task is to show that any sequence $(\bgamma_k) \subset \Gamma^{[a,b]}_{\mN}(\K_1, \K_2)$ has a subsequence convergent in $\Gamma^{[a,b]}_{\mN}$. One can find such a subsequence proceeding analogously as in the proof of Proposition \ref{Ncausality_Prop1} (v). First, notice that for every $j=1,\ldots,N$ one has $(\bgamma^j_k) \subset A^{[a,b]}_\T(\iota^j(\K_1), \iota^j(\K_2))$, where the latter set is compact by the preceding discussion. We can thus pick a sequence of indices $(k_l)$ such that $(\bgamma^1_{k_l})$ converges in $A^{[a,b]}_\T$, then take a subsequence of the sequence $(k_l)$ such that the corresponding subsequence of $(\bgamma^2_{k_l})$ converges as well, and so on. After $N$ such steps we end up with a subsequence $(k_m)$ of indices such that the corresponding subsequences $(\bgamma^j_{k_m})$ converge in $A^{[a,b]}_\T$ for all $j$'s. But by \eqref{Nconvergence}, this means that $(\bgamma_{k_m})$ converges in $\Gamma^{[a,b]}_{\mN}$, as desired.
\end{pf}

The Polish spaces $\Gamma^I_{\mN}$, $A^I_\T$ and $B^I_\T$ possess an important topological property --- they are \emph{locally compact}.

\begin{lemma}
\label{NcurvesProp4}
For any $t \in I$ the evaluation map $\ev_t: \Gamma^I_{\mN} \rightarrow \MN$, $\bgamma \mapsto \bgamma(t)$ is proper.
\end{lemma}
\begin{pf}\textbf{.}
Fix a compact set $\K \subset \MN$ and take any sequence $(\bgamma_k) \subset \ev_t^{-1}(\K)$. The latter set is closed by the continuity of $\ev_t$, and so to prove its compactness it suffices to find a subsequence of $(\bgamma_k)$ convergent in $\Gamma^I_{\mN}$. 

To this end, let $([a_m,b_m])_m$ be an increasing sequence of compact subintervals of $I$ such that $\bigcup_{m=1}^\infty [a_m, b_m] = I$. It is now crucial to observe that, for any $m \in \sN$,
\begin{align*}
(\bgamma_k|_{[a_m,b_m]})_k \subset \Gamma^{[a_m,b_m]}_{\mN}\left( J^-_{\mN}(\K) \cap \itSigma^N_{a_m}, J^+_{\mN}(\K) \cap \itSigma^N_{b_m} \right),
\end{align*}
where the latter set is a compact subset of $\Gamma^{[a_m,b_m]}_{\mN}$ on the strength of Proposition \ref{Ncausality_Prop1} (vi) and Proposition \ref{NcurvesProp3}. Bearing that in mind, we can find a convergent subsequence of $(\bgamma_k)$ using the following version of the standard diagonal argument.

Firstly, let $(\bgamma_{1,k})_k$ be a subsequence of $(\bgamma_k)$ such that $(\bgamma_{1,k}|_{[a_1,b_1]})_k$ converges in $\Gamma^{[a_1,b_1]}_{\mN}$. Then, inductively for $m = 2,3,\ldots$, let $(\bgamma_{m,k})_k$ be a subsequence of $(\bgamma_{m-1,k})_k$ such that $(\bgamma_{m,k}|_{[a_m,b_m]})_k$ converges in $\Gamma^{[a_m,b_m]}_{\mN}$. One now simply notices that the sequence $(\bgamma_{k,k})_k$, i.e. the ``diagonal'' subsequence of the sequence $(\bgamma_k)$, has the property that $(\bgamma_{k,k}|_{[a,b]})_k$ converges in $\Gamma^{[a,b]}_{\mN}$ for any $[a,b] \subset I$. But this is equivalent to saying that $(\bgamma_{k,k})_k$ converges in $\Gamma^I_{\mN}$, as desired.
\end{pf}
\begin{proposition}
\label{NcurvesProp5}
The spaces $\Gamma^I_{\mN}$, $A^I_\T$ and $B^I_\T$ are locally compact.
\end{proposition}
\begin{pf}\textbf{.}
Let $\bgamma \in \Gamma^I_{\mN}$. Pick any $t \in I$ and let $\K$ be some compact neighborhood of $\bgamma(t)$ in $\MN$. Then $\ev_t^{-1}(\K)$ is a neighborhood of $\bgamma$ in $\Gamma^I_{\mN}$, which is compact by Lemma \ref{NcurvesProp4}.

The local compactness of the other two spaces follows from the homeomorphisms $\Gamma^I_{\scriptscriptstyle (1)} \cong A^I_\T \cong B^I_\T$ (cf. the proof of Proposition \ref{NcurvesProp1} above).
\end{pf}

Let us finish this subsection by discussing the question of splitting-independence of the causal curves in case when $I = \sR$. Of course, the causal curves $\gamma \in A^\sR_\T$ are by definition parametrised in accordance with the fixed Cauchy temporal function $\T$ and as such they implicitly depend on the chosen GBS splitting. In order to make them manifestly splitting-independent objects, one has to ``deparametrise'' them, what amounts to passing to their images $\im \gamma \vc \gamma(\sR) \subset \M$. In \cite[Proposition 8]{Miller17a} it was shown\footnote{Note that in \cite{Miller17a} the space $A^\sR_\T$ was called $\I_\T$, whereas $\im \gamma$ was denoted by $[\gamma]$.} that the map $\im: A^\sR_\T \rightarrow \Cf_\textrm{inext}$ is a well-defined bijection on the set of all inextendible unparametrised causal curves in $\M$.

What is more, the above bijection can be used to endow $\Cf_\textrm{inext}$ with the locally compact Polish space topology transported from $A^\sR_\T$. This topology is independent from the particular choice of $\T$, as attested by the following fact.
\begin{proposition} \cite[Proposition 9]{Miller17a}
Let $\T_1,\T_2$ be Cauchy temporal functions on $\M$. Then the map $\widetilde{\ }: A^\sR_{\T_1} \rightarrow A^\sR_{\T_2}$ defined via $\widetilde{\gamma} \vc \gamma \circ (\T_2 \circ \gamma)^{-1}$ is a well-defined reparametrisation of causal curves and a homeomorphism.
\end{proposition}
In this way, the map $\im: A^\sR_\T \rightarrow \Cf_\textrm{inext}$ becomes a homeomorphism. Reasoning similarly as in the proof of Proposition \ref{NcurvesProp1}, we thus obtain the following result.
\begin{corollary}
\label{NcurvesCor} The space $\Gamma^\sR_{\mN}$ is homeomorphic with $\Cf_\textup{inext}^N$, i.e. with the $N$-th Cartesian power of the space of all inextendible unparametrised causal curves in $\M$.
\end{corollary}

It is straightforward to see that the homeomorphism ``deparametrising'' the $N$-particle causal curves is nothing but $\bgamma \mapsto (\im \bgamma^1, \ldots, \im \bgamma^N)$.


\subsection{\texorpdfstring{Evolution of $N$-particle measures}{Evolution of N-particle measures}}

Having established the locally compact Polish space $\Gamma^I_{\mN}$ as the suitable $N$-particle version of the `space of worldlines', we can formulate the $N$-particle analogue of Theorem \ref{PolishThm}. To this end, let us introduce the term \emph{evolution of $N$-particle measures} to denote any measure-valued map $I \ni t \mapsto \bmu_t \in \Pf(\MN)$ such that $\supp \bmu_t \subset \itSigma^N_t$. In other words, for any $t \in I$ we have $\bmu_t = \delta_t \times \mu_t^{\mN}$ for a certain $\mu_t^{\mN} \in \Pf(\itSigma^N)$.

\begin{theorem}
\label{NcurvesMain}
Let $I \ni t \mapsto \bmu_t$ be an evolution of measures. The following conditions are equivalent:
\begin{enumerate}[i)]
\item The evolution $t \mapsto \bmu_t$ is \emph{causal}, by which we mean that
\begin{align}
\label{Ncausal_evo}
\forall \, s,t \in I \quad s \leq t \ \Rightarrow \ \bmu_s \preceq \bmu_t
\end{align}
in the sense of Definition \ref{def:caus}.
\item There exists $\bsigma \in \Pf(\Gamma^I_{\mN})$ such that $(\ev_t)_\sharp \bsigma = \bmu_t$ for every $t \in I$, where $\ev_t: \Gamma^I_{\mN} \rightarrow \MN$ denotes the evaluation map.
\end{enumerate}
\end{theorem}

On the strength of Corollary \ref{NcurvesCor}, for $I = \sR$ the measure $\bsigma$ can be reinterpreted\footnote{Formally, one considers the pushforward $(\im \circ \, \iota^1, \ldots, \im \circ \, \iota^N)_\#\bsigma$.} as an element of $\Pf(\Cf_\textup{inext}^N)$. The latter is an invariant object, independent of any particular choice of the GBS splitting of the underlying globally hyperbolic spacetime $\M$. This is a remarkable fact. It shows that to any causal time-evolution of $N$ particles, modelled with the help of a chosen time-function on $\M$, there corresponds a single global object in the manifestly invariant space $\Pf(\Cf_\textup{inext}^N)$. The time-evolution of $N$ particles described via a map $\sR \ni t \mapsto \bmu_t \in \Pf(\MN)$ is thus indeed generally covariant, although even the $N$-particle configuration spacetime $\MN$ itself depends upon the splitting. Another observer, who employs a different time function $\T'$ on $\M$, witnesses an evolution $s \mapsto \bmu'_s \in \Pf(\MN')$ obtained from $\bsigma$ as $(\ev_s)_\sharp \bsigma = \bmu'_s$ for every $s \in \sR$. It is even more remarkable that general covariance holds in the extended context of probability measures, which incorporate correlations between particles.

In order to prove Theorem \ref{NcurvesMain}, we need to carefully adapt all necessary lemmas from \cite{Miller17a} to the $N$-particle setting.

Before delving in, let us recall that the set $\Pf(\X)$ of all (Borel probability) measures on the Polish space $\X$ is itself Polish when endowed with the \emph{narrow topology}. A sequence $(\mu_k) \subset \Pf(\X)$ converges narrowly to some $\mu \in \Pf(\X)$ iff $\int_\X f d\mu_k \rightarrow \int_\X f d\mu$ for all $f \in C_b(\X)$. For an excellent exposition of measure theory on Polish spaces, the Reader is referred to \cite{garling2017polish}.

The first lemma is a straightforward analogue of \cite[Lemma 8]{Miller17a}.
\begin{lemma}
\label{Nnarrowcompactness}
$\itPi_\preceq(\bmu,\bnu)$ is a narrowly compact subset of $\Pf(\MN^2)$.
\end{lemma}

The second lemma is a variant of \cite[Proposition 10]{Miller17a} suitable for the $N$-particle space of causal curves.
\begin{lemma}
\label{Nproperonto}
For any $a,b \in \sR$ the continuous map
\begin{align*}
(\ev_a, \ev_b): \Gamma^{[a,b]}_{\mN} \rightarrow J^+_{\mN} \cap (\itSigma^N_a \times \itSigma^N_b), \qquad (\ev_a, \ev_b)(\bgamma) \vc (\bgamma(a), \bgamma(b))
\end{align*}
is onto, proper and admitting a Borel right inverse.
\end{lemma}
\begin{pf}\textbf{.}
In order to show surjectiveness, take any $(\bp,\bq) \in J^+_{\mN} \cap (\itSigma^N_a \times \itSigma^N_a)$. We thus have $\pi^0(\bp) = a$, $\pi^0(\bq) = b$ and there exists a causal curve $\bgamma: [0,1] \rightarrow \MN$ connecting them. All we have to do is reparametrise $\bgamma$ so that it becomes an element of $\Gamma^{[a,b]}_{\mN}$. To this end, notice that the map $\pi^0 \circ \bgamma: [0,1] \rightarrow [a,b]$ is a well-defined, continuous and strictly increasing surjection. Therefore, $(\pi^0 \circ \bgamma)^{-1}$ exists and is the desired reparametrisation; $\bgamma \circ (\pi^0 \circ \bgamma)^{-1} \in \Gamma^{[a,b]}_{\mN}$. 

In order to show properness, take any compact $\K \subset J^+_{\mN} \cap (\itSigma^N_a \times \itSigma^N_b)$ and notice that
\begin{align*}
(\ev_a, \ev_b)^{-1}(\K) \subset \ev_a^{-1}(\pi^L(\K)) \cap \ev_b^{-1}(\pi^R(\K)),
\end{align*}
which, on the strength of Lemma \ref{NcurvesProp4} and by the continuity of the projection and evaluation maps, means that $(\ev_a, \ev_b)^{-1}(\K)$ is compact as a closed subset of a compact set.

To prove the remaining part of the lemma's statement we invoke the standard measurable selection result, by which a continuous map from a $\sigma$-compact metrisable space onto a metrisable space admits a Borel right inverse \cite[Corollary I.8]{Fabec}.

The space $\Gamma^{[a,b]}_{\mN}$ is a locally compact Polish space (Propositions \ref{NcurvesProp1} \& \ref{NcurvesProp5}) and hence it is metrisable and $\sigma$-compact. Of course, $J^+_{\mN} \cap (\itSigma^N_a \times \itSigma^N_b)$, being a subset of the manifold $\MN^2$, is metrisable, too. Together with the already proven surjectivity of the map $(\ev_a, \ev_b)$, the above-mentioned measurable selection result completes the proof.
\end{pf}

Consider now two curves $\bgamma_1 \in \Gamma^{[a,b]}_{\mN}$ and $\bgamma_2 \in \Gamma^{[b,c]}_{\mN}$ such that $\bgamma_1(b) = \bgamma_2(b)$. Concatenating them yields a new curve $\bgamma_1 \sqcup \bgamma_2: [a,c] \rightarrow \MN$ which evidently belongs to $\Gamma^{[a,c]}_{\mN}$. We now want, however, to extend the concatenation operation onto measures over the spaces of $N$-particle curves. The following definition mimics \cite[Definition 5]{Miller17a}.
\begin{definition}
\label{Nconcat}
For any fixed $a,b,c \in \sR$, $a<b<c$, define the (Polish) space $\digamma^{[a,b,c]}_{\mN} \vc \{ (\bgamma_1, \bgamma_2) \in \Gamma^{[a,b]}_{\mN} \times \Gamma^{[b,c]}_{\mN} \, | \, \bgamma_1(b) = \bgamma_2(b) \}$ and let $\sqcup: \digamma^{[a,b,c]}_{\mN} \rightarrow \Gamma^{[a,c]}_{\mN}$ be the concatenation map. For any $\bsigma_1 \in \Pf(\Gamma^{[a,b]}_{\mN})$ and $\bsigma_2 \in \Pf(\Gamma^{[b,c]}_{\mN})$ we say they are \emph{concatenable} if $(\ev_b)_\sharp \bsigma_1 = (\ev_b)_\sharp \bsigma_2 \cv \bnu$ and we define their \emph{concatenation} $\bsigma_1 \sqcup \bsigma_2 \in \Pf(\Gamma^{[a,c]}_{\mN})$ with the help of the Riesz--Markov--Kakutani representation theorem via
\begin{align*}
\int_{\Gamma^{[a,c]}_{\mN}} f d(\bsigma_1 \sqcup \bsigma_2) \vc \int_{\itSigma^N_b} \left( \int_{\digamma^{[a,b,c]}_{\mN}} f(\bgamma_1 \sqcup \bgamma_2) \, d(\bsigma^{\bp}_1 \times \bsigma^{\bp}_2) (\bgamma_1,\bgamma_2) \right) d\bnu(\bp),
\end{align*}
for any $f \in C_c(\Gamma^{[a,c]}_{\mN})$, where $\{\bsigma^{\bp}_i \}_{\bp \in \itSigma^N_b}$ is the disintegration of $\bsigma_i$ with respect to the map $\ev_b$ for $i=1,2$.
\end{definition}
Notice that in invoking the Riesz--Markov--Kakutani theorem we rely on the local compactness of the spaces $\Gamma^I_{\mN}$ (Proposition \ref{NcurvesProp5}). Of course, one can similarly define the concatenation of measures in the case where one or both spaces of curves involve noncompact intervals. One can also easily verify that
\begin{align*}
(\ev_t)_\sharp (\bsigma_1 \sqcup \bsigma_2) = \left\{ \begin{array}{ll} (\ev_t)_\sharp \bsigma_1 & \textrm{for } t < b
																	 \\ \bnu & \textrm{for } t = b
																	 \\ (\ev_t)_\sharp \bsigma_2 & \textrm{for } t > b
													  \end{array} \right.
\end{align*}

The last lemma we need states that any causal evolution of $N$-particle measures is narrowly continuous (cf. \cite[Proposition 11]{Miller17a})
\begin{lemma}
\label{Nnarrowcont}
Consider a map $t \mapsto \bmu_t \in \Pf(\MN)$ such that $\supp \bmu_t \subset \itSigma^N_t$. If this map satisfies \eqref{Ncausal_evo}, then it is narrowly continuous.
\end{lemma}
\begin{pf}\textbf{.}
We adapt the proof from \cite{Miller17a}. Fix any $[a,b] \in I$. Let us first show that the family $\{\bmu_t\}_{t \in [a,b]}$ is tight.

Indeed, fix any $\varepsilon > 0$ and take a compact $K_a \subset \itSigma^N_a$ such that $\bmu_a(K_a) \geq 1 - \varepsilon$, which can always be done on the strength of Ulam's tightness theorem. Let $\K \vc J^+_{\mN}(K_a) \cap J^-_{\mN}(\itSigma^N_b)$, which is compact by Proposition \ref{Ncausality_Prop1} (vii). For any $t \in [a,b]$ one has, of course, that $\supp \bmu_t \subset \itSigma^N_t \subset J^-_{\mN}(\itSigma^N_b)$ and thus one can write
\begin{align*}
\bmu_t(\K) = \bmu_t(J^+_{\mN}(K_a) \cap J^-_{\mN}(\itSigma^N_b)) = \bmu_t(J^+_{\mN}(K_a)) \geq \bmu_a(J^+_{\mN}(K_a)) = \bmu_a(K_a) \geq 1 - \varepsilon,
\end{align*}
where we have used condition \eqref{Ncausal_evo} and one of the characterisations of $\preceq$ (condition \eqref{causal2}).

Our aim now is to show that in the narrow topology $\lim_{s \rightarrow 0^+} \bmu_{t+s} = \bmu_t$ for any fixed $t \in [a,b)$. On the strength of \cite[Lemma 7]{Miller17a}, the tightness of $\{\bmu_t\}_{t \in [a,b]}$ allows us to consider only the compactly supported test functions. In other words, it suffices to prove that $\lim_{s \rightarrow 0^+} \int_{\MN} f d\bmu_{t+s} = \int_{\MN} f d\bmu_t$ for all $f \in C_c(\MN)$.

For any $0 < s \leq b-t$ choose $\bomega_{t,s} \in \itPi_\preceq(\bmu_t, \bmu_{t+s})$ existing by \eqref{Ncausal_evo} and observe that
\begin{align}
\label{Nnarrowcont1}
\left| \int_{\MN} f d\bmu_t  - \int_{\MN} f d\bmu_{t+s} \right| \leq \int_{\supp \bomega_{t,s} \setminus (K_f^c)^2} |f(\bp) - f(\bq)| d\bomega_{t,s}(\bp,\bq),
\end{align}
Where $K_f$ denotes the (compact) support of $f$. We will show that the rightmost integral can be made arbitrarily small for $s$ sufficiently close to $0$.

To this end, let us define an auxiliary complete Riemannian metric on $\MN$ as follows. Recall that, by Theorem \ref{GBSthm}, the metric on $\M$ can be expressed as $g = -\alpha d\T \otimes d\T + \bar{g}$. Then $h \vc \alpha d\T \otimes d\T + \bar{g} = g + 2 \alpha d\T \otimes d\T$ is a Riemannian metric on $\M$ and, moreover,
\begin{align*}
w_0 \vc \sum\limits_{j=1}^N (\iota^j)^\ast h = \sum\limits_{j=1}^N (\iota^j)^\ast g + 2 \sum\limits_{j=1}^N (\iota^j)^\ast \alpha \, d\pi^0 \otimes d\pi^0
\end{align*}
is a Riemannian metric on $\MN$. By the Nomizu--Ozeki theorem \cite{NomizuOzeki}, there exists a positive map $u \in C^\infty(\MN)$ such that $w \vc u w_0$ is a complete Riemannian metric on $\MN$.

Let $d_w$ denote the distance function associated with $w$. We claim that
\begin{align}
\label{Nnarrowcont2}
\forall (\bp,\bq) \in \supp \bomega_{t,s} \setminus (K_f^c)^2 \qquad d_w(\bp,\bq) \leq C \cdot s,
\end{align}
where the constant $C$ depends only on $a,b,f$. With the aid of this inequality one can easily bound the rightmost integral in \eqref{Nnarrowcont1} by any $\varepsilon$. Indeed, since $f$ is uniformly continuous (by the Heine--Cantor theorem), there exists $\delta$ such that for all $\bp,\bq \in K_f$ inequality $d_w(\bp,\bq) < \delta$ implies $|f(\bp) - f(\bq)| < \varepsilon$. Thus, for $s < \delta/C$ one would get 
\begin{align*}
\int_{\supp \bomega_{t,s} \setminus (K_f^c)^2} |f(\bp) - f(\bq)| d\bomega_{t,s}(\bp,\bq) \leq \varepsilon \int_{\supp \bomega_{t,s} \setminus (K_f^c)^2} d\bomega_{t,s}(\bp,\bq) \leq \varepsilon.
\end{align*}
In order to prove \eqref{Nnarrowcont2}, observe first that
\begin{align*}
d_w(\bp,\bq) & = d_w(\bgamma(t),\bgamma(t+s)) \leq \int_t^{t+s} \sqrt{w(\bgamma'(\tau), \bgamma'(\tau))} \, d\tau
\\
& = \int_t^{t+s} \sqrt{u(\bgamma(\tau))} \cdot \sqrt{ \sum\nolimits_j g(\bgamma^{j\prime}(\tau), \bgamma^{j\prime}(\tau)) + 2 \sum\nolimits_j \alpha(\bgamma^j(\tau))} \, d\tau
\\
& \leq \int_t^{t+s} \sqrt{2 u(\bgamma(\tau)) \sum\nolimits_j \alpha(\bgamma^j(\tau))} \, d\tau \, \leq \, s \cdot \max\nolimits_{\br \in J^+_{\mN}(\bp) \cap J^-_{\mN}(\bq)} \sqrt{2 u(\br) \sum\nolimits_j \alpha(\br^j)} \, ,
\end{align*}
where a suitable $\bgamma$ exists by the `onto' part of Lemma \ref{Nproperonto}. We have also used the fact that all $\bgamma^j$'s must be causal (cf. Definition \ref{Ncausality_Def}) and so $g(\bgamma^{j\prime}, \bgamma^{j\prime}) \leq 0$.

However, $\max\nolimits_{\br \in J^+_{\mN}(\bp) \cap J^-_{\mN}(\bq)} \sqrt{2 u(\br) \sum\nolimits_j \alpha(\br^j)}$ is not yet the desired constant $C$, because it depends on $\bp$ and $\bq$ (and so on $s$ and $t$ as well) defining the maximization domain. Thus, what we need is a compact superset $\K \supset \supp \bomega_{t,s} \setminus (K_f^c)^2$ that would manifestly depend only on $a,b,f$ and then put
\begin{align*}
C \vc \max\nolimits_{\br \in J^+_{\mN}(\pi^L(\K)) \cap J^-_{\mN}(\pi^R(\K))} \sqrt{2 u(\br) \sum\nolimits_j \alpha(\br^j)} \, .
\end{align*}
One possible example of such a superset is
\begin{align*}
\K \vc K_f \times \left[ J^+_{\mN}(K_f) \cap J^-_{\mN}(\itSigma^N_b) \right] \, \cup \, \left[ J^-_{\mN}(K_f) \cap J^+_{\mN}(\itSigma^N_a) \right] \times K_f,
\end{align*}
the compactness of which follows from Proposition \ref{Ncausality_Prop1} (vii).

This concludes the proof that $\lim_{s \rightarrow 0^+} \bmu_{t+s} = \bmu_t$ for any fixed $t \in [a,b)$. The proof for the other one-sided limit is completely analogous. Because the interval $[a,b] \subset I$ was arbirtary, it follows that the map $t \mapsto \bmu_t$ is narrowly continuous on the whole $I$.
\end{pf}

Having carefully adapted all the necessary tools and lemmas from \cite{Miller17a} to the $N$-particle setting, we are ready to prove Theorem \ref{NcurvesMain}. The line of reasoning is based on the one conducted in \cite{Miller17a}. Let us present how the above definitions and lemmas play their part in the current setting.

\begin{pf}\textbf{ of Theorem \ref{NcurvesMain}}
(ii) $\Rightarrow$ (i): Fix $s,t \in I$, $s < t$. Similarly as in Lemma \ref{Nproperonto}, consider the `pair-evaluation' map $(\ev_s, \ev_t): \Gamma^I_{\mN} \rightarrow \MN^2$ and define $\bomega \vc (\ev_s, \ev_t)_\sharp \bsigma$. We need to show that $\bomega \in \itPi_\preceq(\bmu_s, \bmu_t)$. Indeed, one has that
\begin{align*}
(\pi^L)_\sharp \bomega = [\pi^L \circ (\ev_s, \ev_t)]_\sharp \bsigma = (\ev_s)_\sharp \bsigma = \bmu_s
\end{align*}
and similarly $(\pi^R)_\sharp \bomega = \bmu_t$. Moreover,
\begin{align*}
\bomega(J^+_{\mN}) = \sigma((\ev_s, \ev_t)^{-1}(J^+_{\mN})) = \sigma(\Gamma^I_{\mN}) = 1,
\end{align*}
where we have used the fact that the image of the map $(\ev_s, \ev_t)$ is a subset of $J^+_{\mN}$ (cf. the first part of the proof of Lemma \ref{Nproperonto}).

(i) $\Rightarrow$ (ii). \textbf{Step 1. The $I = [a,b]$ case.} The idea is to construct a sequence $(\bsigma_n) \subset \Pf(\Gamma^{[a,b]}_{\mN})$ such that $(\ev_t)_\sharp \bsigma_n = \bmu_t$ for all $t$ of the form $t^n_i \vc a + i(b-a)/2^n$, $i=0,1,2,3,\ldots,2^n$ and then show that is has a subsequence convergent to some $\bsigma \in \Pf(\Gamma^{[a,b]}_{\mN})$. Thanks to Lemma \ref{Nnarrowcont}, such a $\bsigma$ must in fact satisfy the above equality for all $t \in [a,b]$, as desired.

The sequence can be constructed $(\bsigma_n)$ as follows. For any fixed $n$ and any $i=1,2,3,\ldots,2^n$, let $S^i: J^+_{\mN} \cap (\itSigma^N_{t^n_{i-1}} \times \itSigma^N_{t^n_i}) \rightarrow \Gamma^{[t^n_{i-1},t^n_i]}_{\mN}$ be the Borel inverse of the map $(\ev_{t^n_{i-1}}, \ev_{t^n_i})$, existing by Lemma \ref{Nproperonto}. Furthermore, let $\bomega_i \in \itPi_\preceq(\bmu_{t^n_{i-1}}, \bmu_{t^n_i})$. Notice that each $\bomega_i$ can be regarded as an element of $\Pf(J^+_{\mN} \cap (\itSigma^N_{t^n_{i-1}} \times \itSigma^N_{t^n_i}))$. Using the concatenation introduced in Definition \ref{Nconcat}, we can thus define
\begin{align*}
\bsigma_n \vc S^1_\sharp \bomega_1 \sqcup S^2_\sharp \bomega_2 \sqcup S^3_\sharp \bomega_3 \sqcup \ldots \sqcup S^{2^n}_\sharp \bomega_{2^n} \in \Pf(\Gamma^{[a,b]}_{\mN}).
\end{align*}
One can easily verify that indeed $(\ev_{t^n_i})_\sharp \bsigma_n = \bmu_{t^n_i}$ for every $i = 0,1,\ldots,2^n$. Similarly as in the proof of (ii) $\Rightarrow$ (i), one can also check that $(\ev_a, \ev_b)_\sharp \bsigma_n \in \itPi_\preceq(\bmu_a,\bmu_b)$, and so the constructed sequence $(\bsigma_n) \subset (\ev_a, \ev_b)^{-1}_\sharp(\itPi_\preceq(\bmu_a,\bmu_b))$.

Now comes the crucial observation: by Lemmas \ref{Nnarrowcompactness} and \ref{Nproperonto}, the set $(\ev_a, \ev_b)^{-1}_\sharp(\itPi_\preceq(\bmu_a,\bmu_b))$ is compact, and thus $(\bsigma_n)$ has a convergent subsequence. Its limit $\bsigma$, as already explained, is the desired measure on the space $\Gamma^{[a,b]}_{\mN}$.

\textbf{Step 2. The $I = [0,\infty)$ case.} For any $i = 1,2,3,\ldots$ denote $\X_i \vc \Gamma^{[i-1,i]}_{\mN}$ and construct $\bsigma_i \in \Pf(\X_i)$ satisfying $(\ev_t)_\sharp \bsigma_i = \bmu_t$ for $t \in [i-1,i]$, as explained in Step 1. The idea now is to perform countable concatenation
\begin{align*}
\bsigma \vc \bsigma_1 \sqcup \bsigma_2 \sqcup \bsigma_3 \sqcup \ldots
\end{align*}
which can be rigorously done with the help of the Kolmogorov extension theorem, yielding $\bsigma \in \Gamma^{[0,\infty)}_{\mN}$ with the desired properties. The details are somewhat tedious and technical, but luckily the exposition given in \cite{Miller17a} remains valid, requiring only certain notational modifications and adjustments, namely: changing $\M$ to $\MN$, $\T^{-1}(n)$ to $\itSigma^N_n$, $C_\T^{[0,\infty)}$ to $\Gamma_{\mN}^{[0,\infty)}$ as well as paying extra attention to the usage of boldface Greek letters, which in \cite{Miller17a} have different meaning.

\textbf{Step 3. The $I = \sR$ case.} Construct $\bsigma_+ \in \Gamma^{[0,\infty)}_{\mN}$ as explained in Step 2. One can analogously construct $\bsigma_- \in \Gamma^{(-\infty,0]}_{\mN}$ such that $(\ev_t)_\sharp \bsigma_- = \bmu_t$ for every $t \leq 0$, in a sense performing the countable concatenation from right to left
\begin{align*}
\bsigma_- \vc \ldots \sqcup \bsigma_3 \sqcup \bsigma_2 \sqcup \bsigma_1
\end{align*}
where this time $\bsigma_i \in \Pf(\Gamma^{[-i,-i+1]}_{\mN})$. Then, one might simply define $\bsigma \vc \bsigma_- \sqcup \bsigma_+$ (cf. remarks following Definition \ref{Nconcat}).

\textbf{Step 4. Remaining cases.} Other types of the interval $I$ can be handled by modifying the approaches presented in the earlier steps.
\begin{itemize}
\item For $I = [a,\infty)$ one defines $\X_i \vc \Gamma^{[a+i-1,a+i]}_{\mN}$ and proceeds as in Step 2.
\item For $I = [a,b)$ one defines $\X_i \vc \Gamma^{[b+(a-b)2^{1-i},b+(a-b)2^{-i}]}_{\mN}$ and proceeds as in Step 2.
\item For $I = (-\infty,b]$ or $I = (a,b]$ one modifies the previous two cases analogously as when constructing $\bsigma_-$ in Step 3.
\item Finally, for $I = (a,b)$, $I = (a,\infty)$ or $I = (-\infty,b)$ one concatenates a suitable pair of $\bsigma$'s from earlier cases.
\end{itemize}
\end{pf}


\section{Multi-particle relativistic wave equations}\label{sec:wave_eq}

In this section we present an application of the developed formalism in quantum wave dynamics. Before turning to concrete examples, we firstly establish a general result linking the causal evolution of measures with the continuity equation.

From now on we specialise to the context of $(1+n)$-dimensional Minkowski spacetime $\sM$ with the natural Cauchy temporal function being the projection $\pi^0$. Of course, the associated GBS splitting is now trivial and our `$N$-particle Minkowski configuration spacetime' is simply $\sR \times \sR^{nN} \cv \sM_{\mN}$. Its points are labelled by $(t,x_1,\ldots,x_N) \cv (t,\bx)$, where $x_j \in \sR^n$ for each $j = 1, \ldots, N$ and $\bx \in \sR^{nN}$.

Every evolution of measures $I \ni t \mapsto \bmu_t$ on $\sM_{\mN}$ can be written in the form $\bmu_t = \delta_t \x \mu^{\mN}_t$ with $\mu^{\mN}_t \in \Pf(\sR^{nN})$ for every $t \in I$.

In what follows, we reintroduce the explicit value of the speed of light in vacuum $c$.


\subsection{The continuity equation}

If an evolution of measures $t \mapsto \mu_t \in \Pf(\M)$ satisfies the continuity equation with a subluminal velocity field, then it is causal in the sense of Definition \ref{def:caus}. This connection, sharpening the intuitions voiced by Gerlach, Gromes, Petzold and Rosenthal \cite{Gerlach1969,Gerlach1968,Gromes1970}, was established in \cite[Section II.B]{PRA2017}. Here we show, that it admits a rather straightforward extension to the multi-particle setting. It is important to recognise that both the ``density'', i.e. the measure $\mu^{\mN}_t \in \Pf(\sR^{nN})$, and the multi-velocity field now include the correlations among the particles.

To begin with, let us define the continuity equation in the $N$-particle context, basing on \cite[Definition 4]{PRA2017} (compare also \cite[Definition 1.4.1]{Crippa2007PhD}).
\begin{definition}
Fix a number $T > 0$ and a Borel map
\begin{align*}
\bv: [0,T] \x \sR^{nN} & \rightarrow \sR^{nN}, \qquad (t,\bx) \mapsto \bv_t(\bx) = \big( v^1_t(\bx), \ldots, v^N_t(\bx) \big)
\end{align*}
called the \emph{multi-velocity field}. We say that a measure-valued map $\mu^{\mN}: [0,T] \rightarrow \Pf(\sR^{nN})$, $t \mapsto \mu^{\mN}_t$ satisfies the \emph{continuity equation} with the multi-velocity field $\bv$ if the equation
\begin{align}
\label{multi}
    \partial_t \mu^{\mN}_t + \sum\limits_{j=1}^N \nabla_j \cdot (\mu^{\mN}_t v^j_t) = 0,
\end{align}
where $\nabla_j$ differentiates with respect to $x_j$, holds in the distributional sense, i.e. if
\begin{align}
    \label{multi1}
    \forall \, f \in C_c^{\infty}((0,T) \x \sR^{nN}) \qquad \int_0^T \int_{\sR^{nN}} \Bigg( \partial_t f + \sum\limits_{j=1}^N v^j_t \cdot \nabla_j f \Bigg) d\mu^{\mN}_t dt = 0.
\end{align}
\end{definition}

Just as in the one-particle case, one finds out that the continuity equation entails a causal evolution of measures, provided that every component of the multi-velocity field is subluminal.
\begin{theorem}\label{thm:cont}
Suppose the map $\mu^{\mN}: [0,T] \rightarrow \Pf(\sR^{nN})$, $t \mapsto \mu^{\mN}_t$ satisfies the continuity equation with the multi-velocity field $\bv$ such that
\begin{align}
\label{subluminal}
    \forall \, t \in [0,T] \quad \forall \, \bx \in \sR^{nN} \quad \forall \, j=1,\ldots,N \qquad \norm{ v^j_t(\bx) } \leq c,
\end{align}
where $\norm{.}$ is the standard Euclidean norm. Then the evolution of measures $[0,T] \ni t \mapsto \bmu_t$ defined via $\bmu_t \vc \delta_t \x \mu^{\mN}_t$ is causal.
\end{theorem}
\begin{pf}\textbf{.}
Just like in the proof for the $N=1$ case \cite[Theorem 3]{PRA2017}, we shall heavily rely on the so-called ``superposition principle'' (see \cite[Theorem 3]{Bernard12} or \cite[Theorem 3.2]{ambrosio2008continuity}). In the present setting, the superposition principle guarantees the existence of a measure $\eta \in \Pf(C([0,T], \sR^{nN}))$ such that
\begin{itemize}
\item $\eta$ is concentrated on the (Borel) set $\cR$ of absolutely continuous maps $\bupr: [0,T] \rightarrow \sR^{nN}$ satisfying $\dot{\bupr}(t) = \bv_t(\bupr(t))$ for $t \in [0,T]$ a.e., and hence one can regard $\eta \in \Pf(\cR)$.
\item For every $t \in [0,T]$ $(\widetilde{\ev}_t)_\sharp \eta = \mu^{\mN}_t$, where $\widetilde{\ev}_t: \cR \rightarrow \sR^{nN}$ is the evaluation map\footnote{We added the tilde $\widetilde{\ }$ to avoid confusion with the evaluation map in Theorem \ref{NcurvesMain}, which will be needed shortly.} $\bupr \mapsto \bupr(t) = (r_1(t), \ldots, r_N(t))$.
\end{itemize}
The latter of the above conditions resembles the defining property of the measure $\bsigma$ appearing in Theorem \ref{NcurvesMain} (ii), and in fact the aim of the current proof is to obtain such a $\bsigma$ from $\eta$.

To this end, let us first show that $\cR \subset (B_{\pi^0}^{[0,T]})^N$, i.e. that for every $\bupr \in \cR$ the curves $t \mapsto \Phi^{-1}(t,r_j(t)) = (t,r_j(t))$ are causal for all $j = 1,\ldots,N$. Indeed, by the absolute continuity of $\bupr$ --- and hence of every $r_j$ --- for any $0 \leq s < t \leq T$ we can write that
\begin{align*}
\norm{ r_j(t) - r_j(s) } = \norm{ \int_s^t \dot{r}_j(\tau)d\tau } \leq \int_s^t \norm{ \dot{r}_j(\tau) } d\tau = \int_s^t \norm{ v^j_t(r_j(\tau)) } d\tau \leq c(t-s),
\end{align*}
where in the last inequality we used the subluminality of $v^j$. Now simply observe that in the Minkowski spacetime the inequality $\norm{ r_j(t) - r_j(s) } \leq c(t-s)$ is equivalent to $(s,r_j(s)) \preceq (t,r_j(t))$. Since $s \neq t$, we obtain that $(s,r_j(s)) \prec (t,r_j(t))$, which concludes the proof that $r_j \in B_{\pi^0}^{[0,T]}$.

Let now $H: \cR \rightarrow \Gamma_{\pi^0}^{[0,T]}$ be defined simply as $H(\bupr) = (\id_{[0,T]}, \bupr)$. As such, it is obviously continuous. Observe that for any fixed $t \in [0,T]$ and any $\bupr \in \cR$ one has
\begin{align*}
(\ev_t \circ H)(\bupr) = (t, \widetilde{\ev}_t(\bupr)),
\end{align*}
where $\ev_t: \Gamma_{\pi^0}^{[0,T]} \rightarrow \sM_{\mN}$ is the evaluation map as used in Theorem \ref{NcurvesMain} (ii). When lifted at the level of $\eta$, the above identity becomes
\begin{align*}
(\ev_t \circ H)_\sharp \eta = \delta_t \x (\widetilde{\ev}_t)_\sharp \eta = \delta_t \x \mu_t^{\mN} = \bmu_t,
\end{align*}
and so to finish the proof it suffices to define $\bsigma \vc H_\sharp \eta$ and invoke Theorem \ref{NcurvesMain}.
\end{pf}

Equipped with Theorem \ref{thm:cont} we are ready to demonstrate the causality of the evolution of probability measures in concrete quantum systems.

\subsection{The multi-photon equation}

Let $\mathbf{E}, \mathbf{B}$ be a spacetime-dependent electromagnetic field and let $u \vc \frac{1}{2}( \varepsilon_0 \norm{\textbf{E}}^2 + \frac{1}{\mu_0} \norm{\textbf{B}}^2)$ be the associated energy density (with $c = 1/\sqrt{\varepsilon_0 \mu_0}$). If the total energy $\E \vc \int_{\sR^3} u \, d^3 x$ is finite then $t \mapsto \mu_t = \tfrac{1}{\E} u(t,x) d^3 x$ defines a legitimate evolution of measures on $\sM$.

It is well known \cite{Landau} that Maxwell equations imply that $\mu_t$ satisfies the continuity equation with a velocity field $v = \mathbf{S}/u$, where $\mathbf{S} \vc \tfrac{1}{\mu_0} \mathbf{E} \times \mathbf{B}$ is the Poynting vector. Since $v$ is subluminal \cite[Example 8]{PRA2017}, the evolution of the normalised energy density of the electromagnetic field is always causal.

The local quantity $\tfrac{1}{\E} \, u$ admits a probabilistic interpretation and can be read as the modulus squared of a photon wave function $\psi$ \cite{Birula94,Birula96,Birula96_2}. The latter belongs to the Hilbert space $L^2(\sR^{3},\sC^{6})$, which includes the two helicity states, and enjoys the Schr\"odinger equation
\begin{align}\label{photon}
\ii \hbar \partial_t \psi(t;x) = - \ii \hbar c \left[ \, \begin{matrix} S \cdot \nabla & 0 \\ 0 & - S \cdot \nabla \end{matrix} \, \right] \psi(t;x),
\end{align}
with $S = (S_1, S_2, S_3)$ denoting the vector of generators of rotations for a spin-1 particle:
\begin{align*}
   S_1 \vc \left[ \, \begin{matrix}
                      0 & 0 & 0 \\
                      0 & 0 & -\ii \\
                      0 & \ii & 0
                        \end{matrix} \, \right], \qquad 
   S_2 \vc \left[ \, \begin{matrix}
                      0 & 0 & \ii \\
                      0 & 0 & 0 \\
                      -\ii & 0 & 0
                        \end{matrix} \, \right], \qquad
   S_3 \vc \left[ \, \begin{matrix}
                      0 & -\ii & 0 \\
                      \ii & 0 & 0 \\
                      0 & 0 & 0
                        \end{matrix} \, \right].
\end{align*}

The concept of a photon wave function extends naturally to the context of many particles resulting in a multi-photon wave function, which is fully compatible with the quantum field theoretic viewpoint (cf. \cite{Multiphoton} and references therein).

An $N$-photon wave function can be written in the following form \cite[Eq. (105)]{Multiphoton}:
\begin{align}\label{N_photon}
\Psi(t;\bx) = \sum\limits_{\alpha \in \sN^N} C_\alpha \bigotimes\limits_{j=1}^N \psi_{\alpha_j}(t; x_j),
\end{align}
where $\{\psi_n\}_{n \in \sN}$ is a set of single-photon basis states and the coefficients $C_\alpha$ are symmetric with respect to the interchange of the multi-index components. For every fixed $t$, $\Psi(t;\cdot)$ is an element of the Hilbert space $L^2(\sR^3,\sC^6)^{\otimes N} \cong L^2(\sR^{3N},(\sC^6)^{\otimes N})$. Hence, we are in a position to construct the measures $\bmu_t \vc \delta_t \x \Psi^\dag(t;\bx)\Psi(t;\bx) d^{3N} \bx$ and study the causality of their evolution.

Observe that the measures $\bmu_t$ are symmetric, which reflects the fact that photons are indistinguishable. 

The Schr\"odinger equation for the multi-photon wave function reads:
\begin{align}\label{Nphot_eq}
\ii \hbar \partial_t \Psi(t;\bx) = - \ii \hbar c \sum\limits_{j=1}^N \beta^{(j)} (S^{(j)} \cdot \nabla^{(j)}) \Psi(t;\bx),
\end{align}
where the superscript ${}^{(j)}$ signifies that the given object acts only on the $j$-th tensor component, i.e.
\begin{align*}
    & \nabla^{(j)} \vc \underbrace{\jdn \otimes \ldots \otimes \jdn}_{j - 1} \otimes \ \left[ \, \begin{matrix}
                      \nabla & 0 \\
                      0 & \nabla
                        \end{matrix} \, \right] \, \otimes \underbrace{\jdn \otimes \ldots \otimes \jdn}_{N-j},
    \\
    & S^{(j)} \vc \underbrace{\jdn \otimes \ldots \otimes \jdn}_{j - 1} \otimes \ \left[ \, \begin{matrix}
                      S & 0 \\
                      0 & S
                        \end{matrix} \, \right] \, \otimes \underbrace{\jdn \otimes \ldots \otimes \jdn}_{N-j},
    \\
    & \beta^{(j)} \vc \underbrace{\jdn \otimes \ldots \otimes \jdn}_{j - 1} \otimes \ \left[ \, \begin{matrix}
                      I_3 & 0 \\
                      0 & -I_3
                        \end{matrix} \, \right] \, \otimes \underbrace{\jdn \otimes \ldots \otimes \jdn}_{N-j}
\end{align*}
with $I_3$ denoting the 3-by-3 identity matrix.

Equation \eqref{Nphot_eq} can be rewritten as
\begin{align*}
\partial_t \Psi(t,\bx) & = - c \sum\limits_{j=1}^N \sum\limits_{k=1}^3 \beta^{(j)} S^{(j)}_k \partial^{(j)}_k \Psi(t;\bx),
\end{align*}
where $\partial^{(j)}_k$ denotes the partial derivative with respect to the $k$-th component of $x_j$.

Multiplying this equation by $\Psi^\dag(t; \bx)$, one gets (after suppressing the arguments):
\begin{align*}
\Psi^\dag \partial_t \Psi = - c \sum\limits_{j=1}^N \sum\limits_{k=1}^3 \Psi^\dag \beta^{(j)} S^{(j)}_k \partial^{(j)}_k \Psi,
\end{align*}
which, when added to its conjugate, immediately yields the continuity equation
\begin{align*}
\partial_t \left( \Psi^\dag \Psi \right) + \sum\limits_{j=1}^N \sum\limits_{k=1}^3 \partial^{(j)}_k \left( \Psi^\dag c \beta^{(j)} S^{(j)}_k \Psi \right) = 0.
\end{align*}

It is not difficult to realise that this is indeed an equation of the form (\ref{multi}). One simply has to take
\begin{align}
\label{N_photon_nut}
    d\mu^{\mN}_t \vc \Psi^\dag(t; \bx) \Psi (t; \bx) d^{3N}\bx
\end{align}
and, for any $j=1,\ldots,N$, define $v^j_t \vc (v^{j,1}_t, v^{j,2}_t, v^{j,3}_t)$ via
\begin{align*}
    v^{j,k}_t(\bx) \vc 
  \begin{cases}
            \frac{ \Psi^\dag(t; \bx) c \beta^{(j)} S^{(j)}_k \Psi(t; \bx)}{\Psi^\dag(t; \bx) \Psi(t; \bx)}, & \textrm{ for } (t; \bx) \textrm{ such that } \Psi^\dag(t; \bx) \Psi(t; \bx) \neq 0,
            \\
            0 & \textrm{ otherwise.}
                        \end{cases} 
\end{align*}
One can show that $\|v^j_t(\bx)\| \leq c$ for every $j=1,\ldots,N$, $\bx \in \sR^{3N}$ and for every $t$ from the considered interval. Indeed, it amounts to demonstrating that for any $\bu, \bw \in \sC^3$
\begin{align}
\label{subc}
    \sum\limits_{k=1}^3 \left( \bu^\dag S_k \bu - \bw^\dag S_k \bw \right)^2
                        \leq
                        \left( \norm{\bu}^2 + \norm{\bw}^2 \right)^2,
\end{align}
where (somewhat abusing the notation) we write $\norm{\ba} \vc \sqrt{\ba^\dag \ba}$ for the standard Euclidean norm on $\sC^3$. The crucial step is to rewrite the left-hand side of (\ref{subc}) as $\norm{ \ii \, \overline{\bu} \x \bu - \ii \, \overline{\bw} \x \bw }^2$, where the overline denotes the complex conjugation. One then has that
\begin{align*}
& \sum\limits_{k=1}^3 \left( \bu^\dag S_k \bu - \bw^\dag S_k \bw \right)^2 = \norm{ \ii \, \overline{\bu} \x \bu - \ii \, \overline{\bw} \x \bw }^2 \leq \left( \norm{\overline{\bu} \x \bu} + \norm{\overline{\bw} \x \bw} \right)^2
\\
& = \left( \sqrt{\norm{\bu}^4 - |\bu^T\bu|^2} + \sqrt{\norm{\bw}^4 - |\bw^T\bw|^2} \right)^2 \leq \left( \norm{\bu}^2 + \norm{\bw}^2 \right)^2,
\end{align*}
where we have used the triangle inequality and the identity $\norm{\ba \x \bb}^2 + |\ba^\dag\bb|^2 = \norm{\ba}^2\norm{\bb}^2$.

On the strength of Theorem \ref{thm:cont}, we therefore obtain that the evolution of the measures $\bmu_t \vc \delta_t \x \mu^{\mN}_t$ is causal in the sense of condition (\ref{Ncausal_evo}).

\subsection{The multi-fermion equation}

The photon equation \eqref{photon} is in fact a spin-1 analogue of the (massless) Dirac equation \cite{Birula94,PenroseRindler}. Therefore, one can construct a ``multi-fermion'' wave function along the same lines:
\begin{align}\label{N_Dirac}
\Psi(t;\bx) = \sum\limits_{\alpha \in \sN^N} A_\alpha \bigotimes\limits_{j=1}^N \psi_{\alpha_j}(t; x_j),
\end{align}
where $\{\psi_n\}_{n \in \sN}$ is a set of single-fermion basis states and the coefficients $A_\alpha$ are now antisymmetric with respect to the interchange of the multi-index components.

For every $t$, $\Psi(t;\cdot)$ belongs to the Hilbert space $L^2(\sR^3,\sC^4)^{\otimes N} \cong L^2(\sR^{3N},(\sC^4)^{\otimes N})$. The resulting measure $\bmu_t \vc \delta_t \x \Psi^\dag(t;\bx)\Psi(t;\bx) d^{3N} \bx$ is again symmetric, as Dirac fermions of the same mass are indistinguishable.

The Schr\"odinger equation for $\Psi$ reads:
\begin{align}\label{N_Dirac_eq}
\ii \hbar \partial_t \Psi = - \ii \hbar c \sum\limits_{j=1}^N \sum\limits_{k=1}^3 \gamma^{(j)}_0 \gamma^{(j)}_k \partial^{(j)}_k \Psi + m c^2 \sum\limits_{j=1}^N \gamma^{(j)}_0 \Psi,
\end{align}
where the arguments have been suppressed and where $\gamma^{(j)}_\mu$'s denote the respective gamma matrices acting at the $j$-th tensor component: 
\begin{align*}
    & \gamma^{(j)}_\mu \vc \underbrace{\jdn \otimes \ldots \otimes \jdn}_{j - 1} \otimes \ \gamma^\mu \, \otimes \underbrace{\jdn \otimes \ldots \otimes \jdn}_{N-j}.
\end{align*}
We adopt here the convention that $\gamma^\mu \gamma^\nu + \gamma^\nu \gamma^\mu = -2\eta^{\mu \nu} \jdn$, $(\gamma^0)^\dag = \gamma^0$, $(\gamma^k)^\dag = -\gamma^k$ for $k=1,2,3$, where $\eta^{\mu \nu} = \textrm{diag}(-1,1,1,1)$.

The wave equation \eqref{N_Dirac_eq} could be seen as modelling $N$ non-interacting Dirac fermions of mass $m$. It provides a formal basis for the Dirac--Fock equations employed in atomic physics and quantum chemistry \cite{Dyall2007,Esteban1999,FroeseFischer2016,Grant2007,Levitt2014}. 

The multi-fermion continuity equation can be derived by analogy with the one-particle case. One begins by multiplying the above equation by $-\tfrac{\ii}{\hbar} \Psi^\dag$, obtaining
\begin{align}
\label{mfce1}
\Psi^\dag \partial_t \Psi = - c \sum\limits_{j=1}^N \sum\limits_{k=1}^3  \Psi^\dag \gamma^{(j)}_0 \gamma^{(j)}_k \partial^{(j)}_k \Psi - \frac{\ii m c^2}{\hbar} \sum\limits_{j=1}^N \Psi^\dag \gamma^{(j)}_0 \Psi.
\end{align}
Taking the Hermitian conjugate of the above equation and using the identity $\left(\gamma^\mu\right)^\dag = \gamma^0 \gamma^\mu \gamma^0$, one gets
\begin{align*}
\left(\partial_t \Psi\right)^\dag \Psi = - c \sum\limits_{j=1}^N \sum\limits_{k=1}^3  \left(\partial^{(j)}_k \Psi\right)^\dag \gamma^{(j)}_0 \gamma^{(j)}_k \Psi + \frac{\ii m c^2}{\hbar} \sum\limits_{j=1}^N \Psi^\dag \gamma^{(j)}_0 \Psi = 0,
\end{align*}
which, when added to (\ref{mfce1}) yields
\begin{align*}
\partial_t \left( \Psi^\dag \Psi \right) + \sum\limits_{j=1}^N \sum\limits_{k=1}^3 \partial^{(j)}_k \left( \Psi^\dag c \gamma^{(j)}_0 \gamma^{(j)}_k \Psi \right) = 0.
\end{align*}
The continuity equation thus obtained is again of the form (\ref{multi}) with $\mu^{\mN}_t$ given by formula (\ref{N_photon_nut}) and the multi-velocity field defined via
\begin{align*}
    v^{j,k}_t(\bx) \vc 
  \begin{cases}
            \frac{ \Psi^\dag(t; \bx) c \gamma^{(j)}_0 \gamma^{(j)}_k \Psi(t; \bx)}{\Psi^\dag(t; \bx) \Psi(t; \bx)}, & \textrm{ for } (t; \bx) \textrm{ such that } \Psi^\dag(t; \bx) \Psi(t; \bx) \neq 0,
            \\
            0 & \textrm{ otherwise,}
                        \end{cases} 
\end{align*}
analogously as in the multi-photon case. Also in this case one can show that $\|v^j_t(\bx)\| \leq c$ for all $j=1,\ldots,N$, $\bx \in \sR^{3N}$ and for every $t$. This boils down to verifying that for any $\bz \in \sC^4$
\begin{align*}
    \sum\limits_{k=1}^3 \left( \bz^\dag \gamma^0 \gamma^k \bz \right)^2
                        \leq
                        \left( \bz^\dag \bz \right)^2,
\end{align*}
which can be demonstrated by direct calculations in any chosen basis of gamma matrices. For instance, in the chiral basis
: $\gamma^0 = \left[ \begin{smallmatrix}
0 & I_2 \\
I_2 & 0
\end{smallmatrix}
\right]$, $\gamma^k = \left[ \begin{smallmatrix}
0 & \sigma^k \\
-\sigma^k & 0
\end{smallmatrix}
\right]$, denoting $\bz = [ z^0, z^1, z^2, z^3 ]^T$ we obtain that
\begin{align*}
 \left( \bz^\dag \bz \right)^2 - \sum\limits_{k=1}^3 \left( \bz^\dag \gamma^0 \gamma^k \bz \right)^2 = 4|z^0 \overline{z}^2 + z^1 \overline{z}^3 |^2 \geq 0.
\end{align*}
For an alternative proof employing the 4-vector nature of $\psi^\dag \gamma^0 \gamma^\mu \psi$ the Reader is referred to \cite[Proposition 10]{PRA2017}.

All in all, we thus obtain that the evolution of measures driven by the multi-fermion equation is causal in the sense of condition (\ref{Ncausal_evo}).


\section{Outlook}

In the present work we have provided a rigorous framework to study the joint dynamics of multiple particles from a generally covariant perspective. To this end, we ventured into a relatively poorly explored domain on the verge of Lorentzian geometry and optimal transport. As an application, we have investigated the causal properties of two multi-particle Schr\"odinger equations utilised in quantum optics and atomic physics. The obtained insights open several new avenues, which seem worth exploring.

On the technical side, it is natural to ask whether the relationship between the causal evolution and the continuity equation described by Theorem \ref{thm:cont} extends beyond the Minkowskian setting. This is quite expected, but the proof would probably require going beyond the ``superposition principle'' invoked above, which explicitly assumes the space to be Euclidean. It is the subject of an ongoing work.


On the conceptual side, it is fairly plausible that the assumption on global hyperbolicity of the primary spacetime $\M$ could be relaxed to causal simplicity or even stable causality. In fact, motivated by Theorem \ref{NcurvesMain}, one might go on to \emph{define} a causal evolution of $N$-particle measures as a probability measure on the $N$-th Cartesian power of the space of unparametrised causal curves (endowed with a suitable topology), without employing a global time function whatsoever. What is more, one could consider endowing the underlying spacetime $\M$ with causal relations defined differently, such as the Sorkin--Woolgar relation \cite{Miller18,MinguzziKCausality,MinguzziUtilities,SorkinWoolgar} or those arising in the context of Lorentz--Finsler geometry \cite{LorentzFinsler,Suhr2016}.

On the theoretical side, it is tempting to examine whether the introduced $N$-particle configuration spacetime could be equipped with some kind of `multi-metric' structure. It would provide a new slant on the bimetric theories of gravity \cite{Bimetric}. Such a structure would also be interesting from the viewpoint of the recently established optimal transport formulation of Einstein equations \cite{Suhr18}.

On the applied side, the concept of causal precedence for single-particle measures has proven useful \cite{PRA2020,PRA2017} in assessing the compatibility of quantum (and even ``post-quantum'' in a well-defined sense) dynamics with the structure of a relativistic spacetime. The results obtained in Section \ref{sec:wave_eq} suggest that this criterion can be extended to the multi-particle setting. However, to do so one needs to explicitly take into account the measurement process \cite{PRA2020}. This is also a work in progress. 


\section*{Acknowledgements}
The work of ME and TM was supported by the National Science Centre in Poland under the research grant Sonatina (2017/24/C/ST2/00322). PH and RH acknowledge support by the Foundation for Polish Science through IRAP project co-financed by EU within Smart Growth Operational Programme (contract no. 2018/MAB/5). We thank Andr\'as L\'aszl\'o for numerous valuable discussions.


\bibliographystyle{abbrv}
\bibliography{causality_bib_multi}{}

\begin{thebibliography}{10}

\bibitem{ambrosio2008continuity}
L.~Ambrosio.
\newblock Transport equation and {C}auchy problem for non-smooth vector fields.
\newblock In B.~Dacorogna and P.~Marcellini, editors, {\em Calculus of
  {V}ariations and {N}onlinear {P}artial {D}ifferential {E}quations: {W}ith a
  historical overview by {E}lvira {M}ascolo}, pages 1--41. Springer, Berlin,
  Heidelberg, 2008.

\bibitem{Beem}
J.~Beem, P.~Ehrlich, and K.~Easley.
\newblock {\em {G}lobal {L}orentzian {G}eometry}, volume 202 of {\em
  {M}onographs and {T}extbooks in {P}ure and {A}pplied {M}athematics}.
\newblock CRC Press, Boca Raton, FL, 1996.

\bibitem{BS04}
A.~Bernal and M.~S{\'{a}}nchez.
\newblock Smoothness of time functions and the metric splitting of globally
  hyperbolic spacetimes.
\newblock {\em Commun. Math. Phys.}, 257(1):43--50, 2005.

\bibitem{Bernard12}
P.~Bernard.
\newblock Some remarks on the continuity equation.
\newblock {\em S\'eminaire \'Equations aux d\'eriv\'ees partielles
  (Polytechnique)}, 2008--2009.
\newblock talk:6.

\bibitem{Bertrand2013}
J.~Bertrand and M.~Puel.
\newblock The optimal mass transport problem for relativistic costs.
\newblock {\em Calc. Var.}, 46(1):353--374, 2013.

\bibitem{Bes09}
F.~Besnard.
\newblock A noncommutative view on topology and order.
\newblock {\em J. Geom. Phys.}, 59(7):861--875, 2009.

\bibitem{BesReview}
F.~Besnard.
\newblock Two roads to noncommutative causality.
\newblock {\em J. Phys. Conf. Ser.}, 634(1):012009, 2015.

\bibitem{Birula94}
I.~Bia{\l}ynicki-Birula.
\newblock On the wave function of the photon.
\newblock {\em Acta Phys. Pol. A}, 86(1):97--116, 1994.

\bibitem{Birula96}
I.~Bia{\l}ynicki-Birula.
\newblock The photon wave function.
\newblock In J.~H. Eberly, L.~Mandel, and E.~Wolf, editors, {\em Coherence and
  {Q}uantum {O}ptics {VII}}, pages 313--322. Springer US, Boston, MA, 1996.
\newblock Proceedings of the Seventh Rochester Conference on Coherence and
  Quantum Optics, held at the University of Rochester, June 7--10, 1995.

\bibitem{Birula96_2}
I.~Bia{\l}ynicki-Birula.
\newblock Photon wave function.
\newblock In E.~Wolf, editor, {\em Progress in {O}ptics {XXXVI}}, pages
  245--294. Elsevier, Amsterdam, 1996.

\bibitem{Brenier2003}
Y.~Brenier.
\newblock Extended {M}onge--{K}antorovich theory.
\newblock In {\em Optimal {T}ransportation and {A}pplications: {L}ectures given
  at the {C.I.M.E.} {S}ummer {S}chool, held in {M}artina {F}ranca, {I}taly,
  {S}eptember 2-8, 2001}, pages 91--121. Springer, Berlin, Heidelberg, 2003.

\bibitem{BrenierFrisch2003}
Y.~Brenier, U.~Frisch, M.~H\'{e}non, G.~Loeper, S.~Matarrese, R.~Mohayaee, and
  A.~Sobolevskii.
\newblock Reconstruction of the early {U}niverse as a convex optimization
  problem.
\newblock {\em Mon. Not. R. Astron. Soc.}, 346(2):501--524, 2003.

\bibitem{CavalettiMondino20}
F.~Cavalletti and A.~Mondino.
\newblock Optimal transport in {L}orentzian synthetic spaces, synthetic
  timelike {R}icci curvature lower bounds and applications.
\newblock 2020.

\bibitem{Crippa2007PhD}
G.~Crippa.
\newblock {\em The flow associated to weakly differentiable vector fields}.
\newblock PhD thesis, Scuola Normale Superiore di Pisa, Universit\"{a}t
  Z\"{u}rich, 2012.

\bibitem{Dyall2007}
K.~G. Dyall and K.~F{\ae}gri~Jr.
\newblock {\em Introduction to Relativistic Quantum Chemistry}.
\newblock Oxford University Press, Oxford, 2007.

\bibitem{UNIV2017}
M.~Eckstein.
\newblock The geometry of noncommutative spacetimes.
\newblock {\em Universe}, 3(1):25, 2017.

\bibitem{PRA2020}
M.~Eckstein, P.~Horodecki, T.~Miller, and R.~Horodecki.
\newblock Operational causality in spacetime.
\newblock {\em Phys. Rev. A}, 101:042128, Apr 2020.

\bibitem{PRA2017}
M.~Eckstein and T.~Miller.
\newblock Causal evolution of wave packets.
\newblock {\em Phys. Rev. A}, 95:032106, Mar 2017.

\bibitem{AHP2017}
M.~Eckstein and T.~Miller.
\newblock Causality for nonlocal phenomena.
\newblock {\em Ann. Henri Poincar{\'e}}, 18:3049--3096, 2017.

\bibitem{Esteban1999}
M.~J. Esteban and E.~S{\'e}r{\'e}.
\newblock Solutions of the {D}irac--{F}ock equations for atoms and molecules.
\newblock {\em Commun. Math. Phys.}, 203(3):499--530, 1999.

\bibitem{Fabec}
R.~Fabec.
\newblock {\em Fundamentals of {I}nfinite {D}imensional {R}epresentation
  {T}heory}.
\newblock Monographs and Surveys in Pure and Applied Mathematics. Chapman and
  Hall/CRC, Boca Raton, FL, 2000.

\bibitem{FroeseFischer2016}
C.~F. Fischer, M.~Godefroid, T.~Brage, P.~J{\"o}nsson, and G.~Gaigalas.
\newblock Advanced multiconfiguration methods for complex atoms: {I}.
  {E}nergies and wave functions.
\newblock {\em J. Phys. B}, 49(18):182004, 2016.

\bibitem{CQG2013}
N.~Franco and M.~Eckstein.
\newblock An algebraic formulation of causality for noncommutative geometry.
\newblock {\em Class. Quantum Gravity}, 30(13):135007, 2013.

\bibitem{Frisch2002}
U.~Frisch, S.~Matarrese, R.~Mohayaee, and A.~Sobolevski.
\newblock A reconstruction of the initial conditions of the {U}niverse by
  optimal mass transportation.
\newblock {\em Nature}, 417(6886):260--262, 2002.

\bibitem{Frisch2011}
U.~Frisch, O.~Podvigina, B.~Villone, and V.~Zheligovsky.
\newblock Optimal transport by omni-potential flow and cosmological
  reconstruction.
\newblock {\em J. Math. Phys.}, 53(3):033703, 2012.

\bibitem{Frisch2006}
U.~Frisch and A.~Sobolevskii.
\newblock Application of optimal transport theory to reconstruction of the
  early universe.
\newblock {\em J. Math. Sci.}, 133(4):1539--1542, 2006.

\bibitem{garling2017polish}
D.~J.~H. Garling.
\newblock {\em Analysis on Polish Spaces and an Introduction to Optimal
  Transportation}.
\newblock London Mathematical Society Student Texts. Cambridge University
  Press, Cambridge, 2017.

\bibitem{Gerlach1969}
B.~Gerlach, D.~Gromes, and J.~Petzold.
\newblock Energie und {K}ausalit{\"{a}}t.
\newblock {\em Z. Phys. A}, 221(2):141--157, 1969.

\bibitem{Gerlach1968}
B.~Gerlach, D.~Gromes, J.~Petzold, and P.~Rosenthal.
\newblock {\"{U}}ber kausales {V}erhalten nichtlokaler {G}r{\"{o}}{\ss}en und
  {T}eilchenstruktur in der {F}eldtheorie.
\newblock {\em Z. Phys. A}, 208(4):381--389, 1968.

\bibitem{GerochSplitting}
R.~Geroch.
\newblock Domain of dependence.
\newblock {\em J. Math. Phys.}, 11(2):437--449, 1970.

\bibitem{Grant2007}
I.~P. Grant.
\newblock {\em Relativistic Quantum Theory of Atoms and Molecules: Theory and
  Computation}, volume~40.
\newblock Springer Science \& Business Media, Berlin, Heidelberg, 2007.

\bibitem{Gromes1970}
D.~Gromes.
\newblock On the problem of macrocausality in field theory.
\newblock {\em Z. Phys.}, 236(3):276--287, 1970.

\bibitem{Hegerfeldt1}
G.~C. Hegerfeldt.
\newblock Remark on causality and particle localization.
\newblock {\em Phys. Rev. D}, 10:3320--3321, 1974.

\bibitem{Hegerfeldt1985}
G.~C. Hegerfeldt.
\newblock Violation of causality in relativistic quantum theory?
\newblock {\em Phys. Rev. Lett.}, 54:2395--2398, 1985.

\bibitem{HegerfeldtFermi}
G.~C. Hegerfeldt.
\newblock Causality problems for {F}ermi's two-atom system.
\newblock {\em Phys. Rev. Lett.}, 72(5):596, 1994.

\bibitem{Hegerfeldt2}
G.~C. Hegerfeldt and S.~N.~M. Ruijsenaars.
\newblock Remarks on causality, localization, and spreading of wave packets.
\newblock {\em Phys. Rev. D}, 22:377--384, 1980.

\bibitem{LorentzFinsler}
M.~A. Javaloyes and M.~S\'anchez.
\newblock Finsler metrics and relativistic spacetimes.
\newblock {\em Int. J. Geom. Methods Mod. Phys.}, 11(09):1460032, 2014.

\bibitem{Kantorovich1}
L.~V. Kantorovich.
\newblock On the translocation of masses.
\newblock {\em Dokl. Akad. Nauk SSSR}, 37:227--229, 1942.

\bibitem{Kantorovich2}
L.~V. Kantorovich.
\newblock On a problem of {M}onge.
\newblock {\em Uspekhi Mat. Nauk}, 3:225--226, 1948.

\bibitem{Kell2018}
M.~Kell and S.~Suhr.
\newblock On the existence of dual solutions for {L}orentzian cost functions.
\newblock 2018.

\bibitem{KP67}
E.~H. Kronheimer and R.~Penrose.
\newblock On the structure of causal spaces.
\newblock {\em Math. Proc. Camb. Philos. Soc.}, 63(2):481--501, 1967.

\bibitem{Landau}
L.~Landau and E.~Lifshitz.
\newblock {\em The Classical Theory of Fields}.
\newblock Course of Theoretical Physics. Butterworth--Heinemann, Oxford, 1975.

\bibitem{Levitt2014}
A.~Levitt.
\newblock Solutions of the multiconfiguration {D}irac--{F}ock equations.
\newblock {\em Rev. Math. Phys.}, 26(07):1450014, 2014.

\bibitem{McCann18}
R.~McCann.
\newblock Displacement convexity of {B}oltzmann's entropy characterizes the
  strong energy condition from general relativity.
\newblock 2018.

\bibitem{Miller17}
T.~Miller.
\newblock On the causality and {$K$}-causality between measures.
\newblock {\em Universe}, 3(1), 2017.

\bibitem{Miller17a}
T.~Miller.
\newblock Polish spaces of causal curves.
\newblock {\em J. Geom. Phys.}, 116:295--315, 2017.

\bibitem{Miller18}
T.~Miller.
\newblock Time functions and {$K$}-causality between measures.
\newblock {\em J. Phys. Conf. Ser.}, 968(1):012008, 2018.

\bibitem{MinguzziKCausality}
E.~Minguzzi.
\newblock {$K$}-causality coincides with stable causality.
\newblock {\em Commun. Math. Phys.}, 290(1):239--48, 2009.

\bibitem{MinguzziUtilities}
E.~Minguzzi.
\newblock Time functions as utilities.
\newblock {\em Commun. Math. Phys.}, 298(3):855--868, 2010.

\bibitem{MS08}
E.~Minguzzi and M.~S{\'a}nchez.
\newblock The causal hierarchy of spacetimes.
\newblock In D.~V. Alekseevsky and H.~Baum, editors, {\em Recent {D}evelopments
  in {P}seudo-{R}iemannian {G}eometry, {ESI} {L}ectures in {M}athematics and
  {P}hysics}, pages 299--358. European Mathematical Society Publishing House,
  Z\"urich, 2008.

\bibitem{Suhr18}
A.~Mondino and S.~Suhr.
\newblock An optimal transport formulation of the {E}instein equations of
  general relativity.
\newblock 2018.

\bibitem{Moretti}
V.~Moretti.
\newblock Aspects of noncommutative {L}orentzian geometry for globally
  hyperbolic spacetimes.
\newblock {\em Rev. Math. Phys.}, 15(10):1171--1217, 2003.

\bibitem{NomizuOzeki}
K.~Nomizu and H.~Ozeki.
\newblock The existence of complete {R}iemannian metrics.
\newblock {\em Proc. Am. Math. Soc.}, 12(6):889--891, 1961.

\bibitem{BN83}
B.~O'Neill.
\newblock {\em Semi-{R}iemannian {G}eometry with {A}pplications to
  {R}elativity}.
\newblock Academic Press, Cambridge, MA, 1983.

\bibitem{Penrose1972}
R.~Penrose.
\newblock {\em Techniques of {D}ifferential {T}opology in {R}elativity},
  volume~7 of {\em CBMS--NSF Regional Conference Series in Applied
  Mathematics}.
\newblock SIAM, Philadelphia, PA, 1972.

\bibitem{PenroseRindler}
R.~Penrose and W.~Rindler.
\newblock {\em Spinors and Space-time}, volume~1.
\newblock Cambridge University Press, Cambridge, 1984.

\bibitem{QIandGR}
A.~Peres and D.~R. Terno.
\newblock Quantum information and relativity theory.
\newblock {\em Rev. Mod. Phys.}, 76:93--123, Jan 2004.

\bibitem{HR09}
H.~Ringstr{\"o}m.
\newblock {\em The {C}auchy {P}roblem in {G}eneral {R}elativity}.
\newblock ESI Lectures in Mathematics and Physics. European Mathematical
  Society, Z\"urich, 2009.

\bibitem{Bimetric}
A.~Schmidt-May and M.~von Strauss.
\newblock Recent developments in bimetric theory.
\newblock {\em J. Phys. A}, 49(18):183001, mar 2016.

\bibitem{2photon}
B.~J. Smith and M.~G. Raymer.
\newblock Two-photon wave mechanics.
\newblock {\em Phys. Rev. A}, 74:062104, Dec 2006.

\bibitem{Multiphoton}
B.~J. Smith and M.~G. Raymer.
\newblock Photon wave functions, wave-packet quantization of light, and
  coherence theory.
\newblock {\em New J. Phys.}, 9(11):414, 2007.

\bibitem{SorkinWoolgar}
R.~Sorkin and E.~Woolgar.
\newblock A causal order for spacetimes with {$C^0$} {L}orentzian metrics:
  proof of compactness of the space of causal curves.
\newblock {\em Class. Quantum Gravity}, 13(7):1971--93, 1996.

\bibitem{Suhr2016}
S.~Suhr.
\newblock Theory of optimal transport for {L}orentzian cost functions.
\newblock {\em M\"unst. J. Math.}, 11:13--47, 2018.

\bibitem{CV03}
C.~Villani.
\newblock {\em Topics in {O}ptimal {T}ransportation}.
\newblock Graduate {S}tudies in {M}athematics. American Mathematical Society,
  Providence (R.I.), 2003.

\end{thebibliography}

\end{document}